\documentclass[journal,romanappendices]{IEEEtran}
\usepackage{amssymb}
\usepackage[cmex10]{amsmath}
\usepackage{mathrsfs}
\usepackage{rotating}
\usepackage{overpic}
\usepackage{algorithm}
\usepackage{algorithmic}
\usepackage{multirow}
\usepackage{array}
\usepackage{graphicx}
\usepackage[caption=false]{subfig}
\usepackage{bm}
\usepackage{tikz}
\usepackage{color}
\usepackage{epstopdf}
\usepackage{stfloats}
\usepackage{cite}
\usepackage{psfrag}
\usepackage{acronym}
\usepackage{enumerate}
\usepackage[braket]{qcircuit}
\xyoption{dvips}
\xyoption{ps}

\usepackage{amssymb}
\usepackage{amsmath}

\definecolor{BLUE}{rgb}{0,0,1}

\newtheorem{proposition}{Proposition}

\newtheorem{assumption}{Assumption}

\newcommand{\tr}[1]{{\rm Tr}\left\{#1\right\}}

\newcommand{\diag}[1]{{\rm diag}\left\{#1\right\}}

\acrodef{ciqem}[CIQEM]{channel-inversion-based quantum error mitigation}
\acrodef{cptni}[CPTnI]{completely positive trace-nonincreasing}
\acrodef{cptp}[CPTP]{completely positive trace-preserving}
\acrodef{ggep}[GGEP]{generalized gate error probability}
\acrodef{nisq}[NISQ]{noisy intermediate-scale quantum}
\acrodef{gep}[GEP]{gate error probability}
\acrodef{mse}[MSE]{mean-squared error}
\acrodef{qaoa}[QAOA]{quantum approximate optimization algorithm}
\acrodef{qcp}[QCP]{quantum channel precoder}
\acrodef{qem}[QEM]{quantum error mitigation}
\acrodef{qpr}[QPR]{quasi-probability representation}
\acrodef{qecc}[QECC]{quantum error correction code}
\acrodef{qedc}[QEDC]{quantum error detection code}
\acrodef{ptm}[PTM]{Pauli transfer matrix}
\acrodef{qpt}[QPT]{quantum process tomography}
\acrodef{rmse}[RMSE]{root-mean-square error}
\acrodef{spam}[SPAM]{state preparation and measurement}
\acrodef{dtft}[DTFT]{discrete-time Fourier transform}
\acrodef{fft}[FFT]{fast Fourier transform}
\acrodef{vqa}[VQA]{variational quantum algorithm}

\acrodefplural{gep}[GEPs]{gate error probabilities}
\acrodefplural{qcp}[QCPs]{quantum channel precoders}
\acrodefplural{qecc}[QECCs]{quantum error correction codes}
\acrodefplural{qedc}[QEDCs]{quantum error detection codes}
\acrodefplural{ptm}[PTMs]{Pauli transfer matrices}
\acrodefplural{vqa}[VQAs]{variational quantum algorithms}

\usepackage{winsnotation}
\usepackage{makecell}
\newcommand{\Sop}[1]{{\mathcal{#1}}}

\definecolor{sotonMarineBlue}{RGB}{1,67,89}
\begin{document}
\title{\huge The Accuracy vs. Sampling Overhead Trade-off in Quantum Error Mitigation Using Monte Carlo-Based Channel Inversion}

\author{Yifeng Xiong, \IEEEmembership{Student Member, IEEE}, Soon Xin Ng, \IEEEmembership{Senior Member, IEEE}, and Lajos Hanzo, \IEEEmembership{Fellow, IEEE}
\thanks{The authors are with School of Electronics and Computer Science, University of Southampton, SO17 1BJ, Southampton (UK).}
\thanks{L. Hanzo would like to acknowledge the financial support of the Engineering and Physical Sciences Research Council projects EP/P034284/1 and EP/P003990/1 (COALESCE) as well as of the European Research Council's Advanced Fellow Grant QuantCom (Grant No. 789028). This work is also supported in part by China Scholarship Council (CSC).}
}

\maketitle

\begin{abstract}
Quantum error mitigation (QEM) is a class of promising techniques for reducing the computational error of variational quantum algorithms. In general, the computational error reduction comes at the cost of a sampling overhead due to the variance-boosting effect caused by the channel inversion operation, which ultimately limits the applicability of QEM.  Existing sampling overhead analysis of QEM typically assumes exact channel inversion, which is unrealistic in practical scenarios. In this treatise, we consider a practical channel inversion strategy based on Monte Carlo sampling, which introduces additional computational error that in turn may be eliminated at the cost of an extra sampling overhead. In particular, we show that when the computational error is small compared to the dynamic range of the error-free results, it scales with the square root of the number of gates. By contrast, the error exhibits a linear scaling with the number of gates in the absence of QEM under the same assumptions. Hence, the error scaling of QEM remains to be preferable even without the extra sampling overhead.  Our analytical results are accompanied by numerical examples.
\end{abstract}

\section*{Acronyms}
\begin{tabular}{ll}
MSE & Mean-Square Error \\
PTM & Pauli Transfer Matrix \\
QEM & Quantum Error Mitigation \\
RMSE & Root-Mean-Square Error \\
VQA & Variational Quantum Algorithm \\
\end{tabular}

\section*{Notations}
\begin{itemize}
\item Deterministic scalars, vectors and matrices are represented by $x$, $\V{x}$, and $\M{X}$, respectively, whereas their random counterparts are denoted as $\rv{x}$, $\RV{x}$, and $\RM{X}$, respectively. Deterministic sets, random sets, and operators are denoted as $\Set{X}$, $\RS{X}$, and $\Sop{X}$, respectively.
\item The notations $\V{1}_n$, $\V{0}_{n}$, $\V{0}_{m\times n}$, and $\M{I}_{k}$, represent the $n$-dimensional all-one vector, the $n$-dimensional all-zero vector, the $m\times n$ dimensional all-zero matrix, and the $k\times k$ identity matrix, respectively. The subscripts may be omitted if they are clear from the context.
\item The notation $\|\V{x}\|_p$ represents the $\ell_p$-norm of vector $\V{x}$, and the subscript may be omitted when $p=2$. For matrices, $\|\M{A}\|_p$ denotes the matrix norm induced by the corresponding $\ell_p$ vector norm. The notation $1/\V{x}$ represents the element-wise reciprocal of vector $\V{x}$.
\item The notation $[\M{A}]_{i,j}$ denotes the $(i,j)$-th entry of matrix $\V{A}$. For a vector $\V{x}$, $[\V{x}]_i$ denotes its $i$-th element. The submatrix obtained by extracting the $i_1$-th to $i_2$-th rows and the $j_1$-th to $j_2$-th columns from $\M{A}$ is denoted as $[\M{A}]_{i_1:i_2,j_1:j_2}$. The notation $[\M{A}]_{:,i}$ represents the $i$-th column of $\M{A}$, and $[\M{A}]_{i,:}$ denotes the $i$-th row, respectively.
\item The trace of matrix $\M{A}$ is denoted as ${\mathrm{Tr}}\{\M{A}\}$, and the complex conjugate of $\M{A}$ is denoted as $\M{A}^\dagger$. Similarly, the complex adjoint of an operator $\Sop{X}$ is also denoted as $\Sop{X}^\dagger$.
\item The notation $\M{A}\otimes \M{B}$ represents the Kronecker product between matrices $\M{A}$ and $\M{B}$. The notation $\M{A}\odot\M{B}$ denotes the Hadamard product between matrices $\M{A}$ and $\M{B}$.
\item The notations $\mathbb{E}\{\cdot\}$, $\mathrm{Var}\{\cdot\}$, and $\mathrm{Cov}\{\cdot\}$ represent the expectation, the variance, and the covariance matrix of their arguments, respectively.
\end{itemize}

\section{Introduction}
\IEEEPARstart{N}{oisy} intermediate-scale quantum computers \cite{nisq} has been one of the most impressive recent advances in the area of quantum computing. In particular, a quantum computer consisting of 53 quantum bits (qubits) has been built in 2019, and has been shown being capable of performing computational tasks that are challenging to be carried out by state-of-the-art classical supercomputers \cite{quantum_supremacy}.

Due to their limited number of qubits, noisy intermediate-scale quantum computers may not be capable of supporting fully fault-tolerant quantum operations relying on quantum error correction codes \cite{qecc,qit,qecc_survey,qtecc}, which are widely believed to be necessary for implementing complex algorithms that require relatively long coherence time \cite{transversal,threshold_thm,fault_tolerance}, such as Shor's factorization algorithm \cite{shor} and Grover's search algorithm \cite{grover}. Alternatively, a class of algorithms tailored for these computers, namely that of the \acp{vqa} \cite{vqe,vqe_theory,qaoa,vqlinear}, is receiving much attention. Briefly, \acp{vqa} aim for sharing their computational tasks between relatively simple quantum circuits and classical computers. A little more specifically, quantum circuits are employed in \ac{vqa} for computing a cost function or its gradient \cite{sgd_vqa}, which is then fed into an optimization algorithm run on classical computers. The objective of this design paradigm is to assist near-term quantum devices in outperforming classical computers in the context of practical problems, such as solving combinatorial optimization problems using the quantum approximate optimization algorithm \cite{qaoa,ansatz,performance_qaoa} and quantum chemistry problems using the variational quantum eigensolver \cite{vqe}.

{
\linespread{1.2}
\begin{table*}[t]
\centering
\footnotesize
\caption{Comparison between the contributions of this treatise and existing literature evaluating the performance of VQAs and QEM.}
\footnotesize
\label{tbl:scaling}
\footnotesize
\begin{tabular}{|l|c|c|c|r|}
\hline
\multirow{2}{*}{} &
\multicolumn{2}{c|}{\textbf{Circuit condition}} &
\multirow{2}{*}{\textbf{Subject of Analysis}} & \multirow{2}{*}{\makecell[r]{\textbf{Method of performance}\\ \textbf{evaluation}}} \\
\cline{2-3}
  & Noisy? & QEM implementation & & \\
\hline
\hline
J. R. McClean \textit{et al.} \cite{vqe_theory} & $\times$ & No QEM & \multirow{3}{*}{Only accuracy} & Only numerical \\
\cline{1-3}\cline{5-5}
 J. R. McClean \textit{et al.} \cite{barren_plateau} & $\times$ & No QEM&  & Analytical and numerical \\
\cline{1-3}\cline{5-5}
S. Wang \textit{et al.} \cite{NIBP} & \checkmark & No QEM &  & Analytical and numerical \\
\hline
S. Endo \textit{et al.} \cite{practical_qem} & \checkmark & Exact channel inversion & \multirow{5}{*}{\makecell[c]{Sampling overhead vs. \\ accuracy trade-off}} & Only numerical \\
\cline{1-3}\cline{5-5}
Y. Xiong \textit{et al.} \cite{sof_analysis} & \checkmark & Exact channel inversion &  & Analytical and numerical \\
\cline{1-3}\cline{5-5}
R. Takagi \cite{ryuji_cost_qem} & \checkmark & Exact channel inversion & & Analytical and numerical \\
\cline{1-3}\cline{5-5}
\textbf{Our contributions} & \checkmark & \makecell[c]{\textbf{Monte Carlo-based} \\ \textbf{channel inversion}} & & Analytical and numerical \\
\hline
\end{tabular}
\end{table*}
}

Although the performance of \acp{vqa} has been characterized using some illustrative examples \cite{vqe,vqe_nearterm,qaoa_decoding,scalable_simulation}, it is not known whether these examples could be scaled up to problems of larger size. In fact, recent analytical results in \cite{barren_plateau,bp2,bp3} support the opposite statement. More explictly, \cite{barren_plateau} proves that the magnitude of the cost function (or its gradient) computed by \acp{vqa} vanishes exponentially as the number of qubits $n$ increases. Fortunately, the follow-up investigations \cite{resort1,resort2} found that this so-called ``barren plateau'' phenomenon may be mitigated to a certain extent by techniques borrowed from the literature of classical machine learning, such as pre-training and layer-by-layer training. However, the authors of \cite{NIBP} show that when decoherence is taken into account, the dynamic range of the computational results also vanishes exponentially upon increasing the circuit depth $N_{\rm L}$, even if these techniques are applied. To summarize, these results imply that when the quantum circuit is long in depth or large in the number of qubits, the computational error become excessive in practical applications.

To improve the error scaling of \acp{vqa} with respect to the depth of the circuits, a body of literature has been devoted to searching for methods that efficiently mitigate the effect of decoherence-induced impairment, without using quantum error correction codes \cite{qem,subspace_expansion1,subspace_expansion2,symmetry_verification}. Among these research efforts, one of the most promising methods is \ac{qem} \cite{qem}, which aims for applying an ``inverse channel'' right after each quantum channel modelling the impact of decoherence. Both the numerical and experimental results of \cite{practical_qem,qem_exp} show that \ac{qem} is indeed capable of reducing the computational error in \acp{vqa} in the context of quantum chemistry problems.

The error reduction capability of \ac{qem} comes at the price of a computational overhead. To elaborate, \ac{qem} is implemented by sampling from a ``quasi-probability representation'' of the inverse channel, which would increase the variance of the computational results, hence some computational overhead (termed as ``sampling overhead'' \cite{practical_qem,sof_analysis}) is required for ensuring that a satisfactory accuracy can be achieved. By appropriately choosing the total number of samples, one may strike a beneficial computational accuracy vs. overhead trade-off. Therefore, it is important to quantify the sampling overhead, before we can conclude whether \ac{qem} can play a significant role in making \acp{vqa} practical.

The literature of \ac{qem} sampling overhead analysis typically assumes that the channel inversion procedure is implemented exactly \cite{practical_qem,ryuji_cost_qem,sof_analysis}.\footnote{We will refer to \ac{qem} based on exact channel inversion as ``ideal \ac{qem}'' in the rest of this treatise.} Under this assumption, the sampling overhead can be characterized by the sampling overhead factor \cite{sof_analysis}, which is determined by the quality of the channel as well as the by basis operations implementing the channel inversion. However, exact channel inversion may be unrealistic in practical scenarios, since it requires a pre-processing stage that is computationally excessive. Moreover, the computational cost of this pre-processing stage increases rapidly with the number of gates, which may negate the benefit of \ac{qem}.

Against this background, in this treatise, we consider a practical channel inversion method based on Monte Carlo sampling, which only increases the pre-processing complexity linearly with the number of gates. The drawback of this method is that it cannot invert the channel exactly, hence there would be some residual error that accumulates during computation. Compared to the ideal \ac{qem}, this method has a less beneficial accuracy vs. overhead trade-off, because additional samples would be necessary to compensate for the residual error. To characterize this trade-off, we investigate the relationship between the residual error and the number of gates $N_{\rm G}$, the number of samples $N_{\rm s}$, and the gate error probability $\epsilon$. We boldly and explicitly contrast our contributions to the related recent research on \acp{vqa} and \ac{qem} in Table \ref{tbl:scaling}, which are further detailed as follows.

\begin{itemize}
\item We analyse the error scaling in the absence of \ac{qem}, by providing both upper and lower bounds of the magnitude of the computation error. We show that the error magnitude scales linearly with the number of gates $N_{\rm G}$, as well as with the gate error probability $\epsilon$, when we have $\epsilon N_{\rm G}\ll 1$.
\item We propose an upper bound on the \ac{rmse} of the computational error in the presence of Monte Carlo-based \ac{qem}. Specifically, we show that the \ac{rmse} is upper bounded by the square root of $N_{\rm G}$ as well as $\epsilon$, when $\epsilon N_{\rm G}\ll 1$.  This implies that when we use the same number of samples as the ideal \ac{qem}, Monte Carlo-based \ac{qem} can still provide a quadratic error reduction versus $N_{\rm G}$, compared to the case of no \ac{qem}.
\item We provide an intuitive interpretation of the proposed error scaling laws, by visualizing the decoherence-induced impairments on the Bloch sphere as the quantum circuit executes.
\item We demonstrate the analytical results using various numerical examples. Specifically, we consider a practical application of carrying out multiuser detection in wireless communication systems using the quantum approximate optimization algorithm and show that our analytical results do apply.
\end{itemize}

The rest of this treatise is organized as follows. In Section \ref{sec:model}, we present the formulation of \acp{vqa} and the channels modelling the decoherence. Then, in Section \ref{sec:qem_implementation}, we discuss a pair of \ac{qem} implementation strategies, namely the Monte Carlo-based \ac{qem} and the exact channel inversion. Based on this discussion, in Section \ref{sec:analysis}, we analyse the error scaling behaviours of these two \ac{qem} implementations, respectively, under the assumption that they use the same number of circuit executions. We provide further intuitions concerning these analytical results in Section \ref{sec:discussion}, with an emphasis on the accuracy vs. sampling overhead trade-off, complemented by numerical examples in Section \ref{sec:numerical}. Finally, we conclude in Section \ref{sec:conclusions}.

\section{Formulation of Variational Quantum Algorithms}\label{sec:model}
\begin{figure*}[t]
\centering
\centering
\begin{overpic}[width=0.8\textwidth]{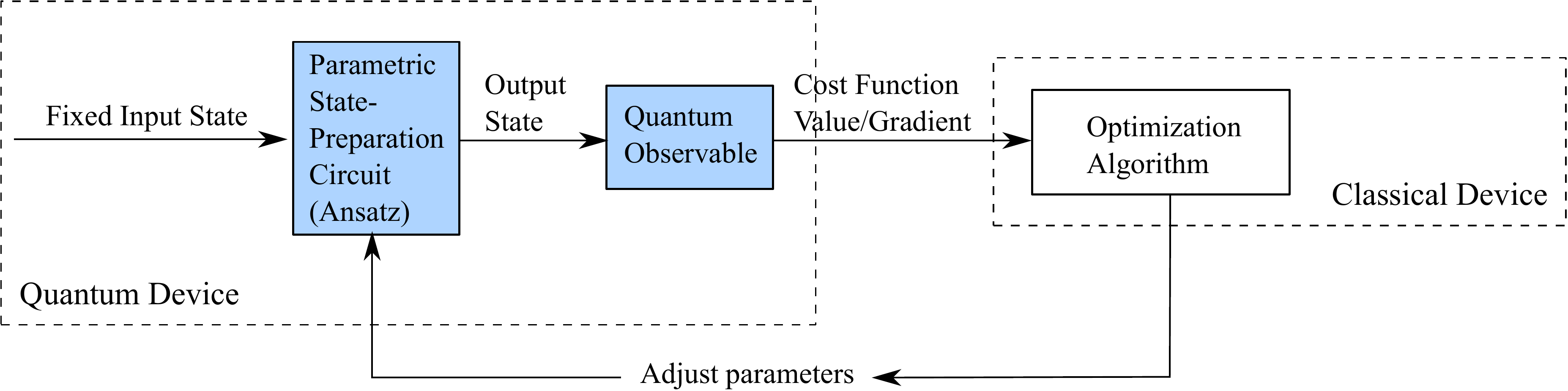}
\end{overpic}
\caption{The structure of a typical implementation for variational quantum algorithms.}
\label{fig:vqa}
\end{figure*}

A typical \ac{vqa} iterates between classical and quantum devices, as portrayed in Fig. \ref{fig:vqa}. The parametric state-preparation circuit (also known as the ``ansatz'' \cite{ansatz}) transforms a fixed input state to an output state, according to the parameters chosen by a classical optimizer. The output state is then measured and fed into a quantum observable, which maps the measurement outcomes to the desired computational results. The results correspond to the value of a cost function or its gradient, which in turn serve as the input of the associated classical optimization algorithm. The iterations continue until certain stopping criterion is met, for example, the computed gradient becomes almost zero.

\begin{figure}[t]
\centering
\begin{overpic}[width=0.49\textwidth]{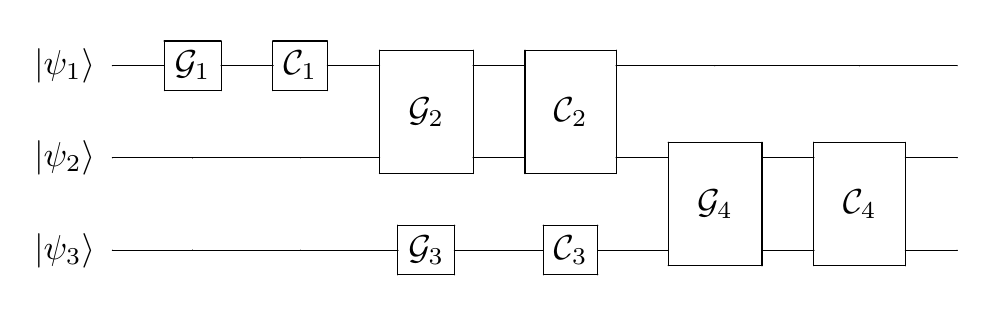}
\end{overpic}
\caption{Simple example of the noisy parametric state-preparation circuit seen in Fig. \ref{fig:vqa}.}
\label{fig:circuit_model}
\end{figure}

In this treatise, we focus on the error induced by the sampling procedure in \ac{qem}, hence we consider the computational result of a single iteration, meaning that the parameters used for state preparation are fixed. We model the decoherence-induced impairment in the parametric state-preparation circuit as quantum channels acting upon the associated quantum states at the output of perfect quantum gates, as exemplified by the simple circuit shown in Fig. \ref{fig:circuit_model}. In this figure, $\Sop{C}_k,~k=1\dotsc 4$ represents the channel modelling the decoherence in the $k$-th quantum gate, while $\Sop{G}_k$ represents the $k$-th ideal decoherence-free quantum gate.

In the subsequent subsections, we present the mathematical formulations of the system models shown in Fig. \ref{fig:vqa} and \ref{fig:circuit_model}.

\subsection{Operator-sum Representation}
Without loss of generality, we assume that the input state of the circuit is the all-zero state $\ket{0}^{\otimes n}$, where $n$ is the number of qubits. In general, when the circuit is decoherence-free, the computational result of a variational quantum circuit may be expressed as
\begin{equation}
\widetilde{r} = \bra{0}^{\otimes n} \left(\prod_{k=1}^{N_{\rm G}}\Sop{G}_i^\dagger\right)\Sop{M}_{\rm ob}\left(\prod_{k=1}^{N_{\rm G}} \Sop{G}_{N_{\rm G}-k+1}\right)\ket{0}^{\otimes n},
\end{equation}
where $N_{\rm G}$ is the number of gates in the circuit, $\Sop{G}_k$ denotes the $k$-th quantum gate, and the operator $\Sop{M}_{\rm ob}$ represents the quantum observable, which describes the computational task as a linear function of the final state.

If we consider a more practical scenario, where the quantum state evolves owing to quantum decoherence as the circuit operates, the state can no longer be fully characterized using the state vector formalism. Instead, we may use the density matrix formalism. In particular, the input state may be described as
\begin{equation}
\rho_0 = (\ket{0} \bra{0})^{\otimes n}.
\end{equation}
Correspondingly, the output state of the $k$-th imperfect quantum gate may be represented in an operator-sum form \cite[Sec. 8.2.4]{ncbook}, relying on following recursive relationship
\begin{equation}\label{recursive_os}
\rho_{k} = \Sop{C}_k\left(\Sop{G}_k\rho_{k-1}\Sop{G}_k^{\dagger}\right),
\end{equation}
where the operator $\Sop{C}_k$ is characterized by
\begin{equation}
\Sop{C}_k(\rho) = \sum_{i=1}^{n_k}\left(\M{E}_{k,i}\rho\M{E}_{k,i}^{\dagger}\right),
\end{equation}
representing the channel modelling the imperfection of the $k$-th gate. The matrices $\M{E}_{k,i}$ represent the operation elements \cite[Sec. 8.2.4]{ncbook} of the channel $\Sop{C}_k$ satisfying the completeness condition of $\sum_{i=1}^{n_k} \M{E}_{k,i}^\dagger \M{E}_{k,i} = \M{I}$. Finally, when all gates completed their tasks and the measurement results have been obtained, the computational result may be expressed as
\begin{equation}\label{computational_result}
r = \tr{\Sop{M}_{\rm ob}\rho_{N_{\rm G}}}.
\end{equation}

\subsection{Pauli Transfer Matrix Representation}
In the standard operator-sum form \cite[Sec. 8.2.4]{ncbook}, the quantum states are represented by matrices. However, in many applications, such as the error analysis considered in this treatise, it would be more convenient to treat them as vectors. Correspondingly, the quantum channels and gates would then be represented by matrices. To this end, the \ac{ptm} representation of quantum operators was proposed in \cite{ptm}, which allows a quantum operator $\Sop{O}$ to be expressed as
\begin{equation}
[\M{O}]_{i,j} = \frac{1}{2^n} \tr{\Sop{S}_i\Sop{O}(\Sop{S}_j)},
\end{equation}
where $\Sop{S}_i$ denotes the $i$-th Pauli operator in the $n$-qubit Pauli group. Similarly, a quantum state $\rho$ can be expressed as
\begin{equation}\label{ptm_vector}
[\V{\rho}]_i = \frac{1}{\sqrt{2^n}} \tr{\Sop{S}_i\rho}.
\end{equation}
Under the \ac{ptm} representation, the computational result may be rewritten as
\begin{equation}\label{ptm_result}
r = \V{v}_{\rm ob}^{\rm T} \left(\prod_{k=1}^{N_{\rm G}} (\M{C}_{N_{\rm G}-k+1}\M{G}_{N_{\rm G}-k+1})\right) \V{v}_0,
\end{equation}
where $\M{G}_k$ represents the $k$-th perfect gate, and $\M{C}_k$ represents the channel modelling the imperfection of the $k$-th gate. The vector $\V{v}_0$ denotes the initial state, whereas $\V{v}_{\rm ob}$ is the vector representation of the quantum observable $\Sop{M}_{\rm ob}$.

To simplify the notation, we define
\begin{equation}
\begin{aligned}
\M{R}_k &:= \prod_{i=1}^k (\M{C}_{k-i+1}\M{G}_{k-i+1}), \\
\widetilde{\M{R}}_k &:= \prod_{i=1}^k \M{G}_{k-i+1}.
\end{aligned}
\end{equation}
Especially, for $k=0$, we define $\M{R}_0=\widetilde{\M{R}}_0 = \M{I}$. The output state of the $k$-th quantum gate can then be expressed as
\begin{equation}\label{ptm_recursion}
\begin{aligned}
\V{v}_k :&= \M{R}_k \V{v}_0 \\
&=\M{C}_k\M{G}_k\V{v}_{k-1}.
\end{aligned}
\end{equation}
Hence we have
\begin{equation}\label{r_in_rng}
\begin{aligned}
r &= \V{v}_{\rm ob}^{\rm T} \V{v}_{N_{\rm G}} \\
&= \V{v}_{\rm ob}^{\rm T}\M{R}_{N_{\rm G}}\V{v}_0.
\end{aligned}
\end{equation}

\subsection{Channel Model}
In this treatise, we consider Pauli channels \cite{pauli_channel}, for which the Pauli transfer matrices take the following form
\begin{equation}\label{pauli_as_diagonal}
\M{C}_k = \diag{\V{c}_k},
\end{equation}
where
\begin{equation}\label{ck_hadamard}
\V{c}_k=\widetilde{\M{H}}\V{p}_k,
\end{equation}
with $\widetilde{\M{H}}$ denoting the Hadamard transform, whereas $\V{p}_k$ represents a probability distribution satisfying $\V{1}^{\rm T}\V{p}_k=1$, $\V{p}_k\ge 0$.

\section{QEM and Its Implementation Strategies}\label{sec:qem_implementation}
Ideally, for a channel $\M{C}_k$, \ac{qem} would apply its inverse based on a linear combination of predefined quantum operations, taking the following form
\begin{equation}
\M{C}_k^{-1} = \sum_{l=1}^L \alpha_l^{(k)} \M{O}_l,
\end{equation}
where $\M{O}_l$ is the $l$-th quantum operation, while $\V{\alpha}_k:=[\alpha_1~\dotsc~\alpha_L]^{\rm T}$ is the quasi-probability representation vector satisfying $\V{1}^{\rm T}\V{\alpha}_k=1$. This linear combination may be rewritten as a probabilistic mixture of the quantum operations as follows:
\begin{equation}\label{inverse_channel}
\M{C}_k^{-1} = \|\V{\alpha}_k\|_1 \sum_{l=1}^L s_l^{(k)} p_l^{(k)} \M{O}_l,
\end{equation}
where $s_l^{(k)}$ and $p_l^{(k)}$ are the $l$-th entries of $\V{s}_k$ and $\V{p}_k$, respectively, given by
\begin{equation}
\begin{aligned}
p_i^{(k)} &= \frac{|\alpha_i^{(k)}|}{\|\V{\alpha}_k\|_1} , \\
\V{s}_k &= \mathrm{sgn}\{\V{\alpha}_k\}.
\end{aligned}
\end{equation}
Note that the vector $\V{p}_k$ describes a probability distribution.

Typically, the probabilistic mixture in \eqref{inverse_channel} is implemented by generating a set of candidate circuits and performing post-processing on the output of these circuits. In the following subsections, we will discuss different candidate selection strategies and their characteristics.

\subsection{Exact Implementation and Sampling Overhead}
The inverse channel $\M{C}_k^{-1}$ in \eqref{inverse_channel} is assumed to be implemented exactly in the seminal paper \cite{qem} that proposed \ac{qem} for the first time, as well as in many other existing contributions \cite{practical_qem,ryuji_cost_qem,sof_analysis}. Exact implementation implies that, each quantum operation $\M{O}_l$ should appear in exactly $Np_l^{(k)}$ candidate circuits in every $N$ samples of the computational result.

The assumption of exact implementation significantly simplifies the performance analysis of \ac{qem}. In particular, it leads to a clear and concise formula of sampling overhead, which describes the computational overhead imposed by the variance-boosting effect of \ac{qem}. To elaborate, assume that the variance of the computational result is $\sigma^2$ based on $N_0$ samples. According to \eqref{inverse_channel}, if we implement the inverse channel $\M{C}_k^{-1}$, the variance would become $\|\V{\alpha}_k\|_1^2\sigma^2$. Therefore, in order to achieve the same accuracy as the case without \ac{qem}, we should acquire $N_0(\|\V{\alpha}_k\|_1^2-1)$ additional samples. If we further assume that all gates are protected by \ac{qem}, we have the following formula for the total sampling overhead
\begin{equation}\label{additional_exact}
N_{\rm exact}=N_0\left(\prod_{k=1}^{N_{\rm G}} \|\V{\alpha}_k\|_1^2-1\right).
\end{equation}
The simplicity of \eqref{additional_exact} is largely due to the assumption of exact implementation.

Despite its theoretical convenience, the practicality of exact implementation is doubtful. Specifically, the number of the $l$-th candidate circuit, $Np_l^{(k)}$ has to be an integer, which might be unrealistic for an arbitrary $p_l^{(k)}$. Furthermore, the number of the probability parameters $p_l^{(k)}$ would increase exponentially as the number of \ac{qem}-protected gates increases, which may render the candidate circuit selection procedure computationally prohibitive when $N_{\rm G}$ is large. Motivated by these drawbacks, we propose to use a Monte Carlo implementation of \ac{qem}, detailed in the next subsection.

\subsection{Monte Carlo Implementation}
In the Monte Carlo implementation, we first sample from the probability distribution $\V{p}_k$ for each gate, and obtain $N$ samples constituting a set $\RS{L}=\{\rv{l}_1,\dotsc,\rv{l}_{N}\}$, where for all $k$ we have $\rv{l}_k=1,2,\dotsc,L$. Thus we may approximate the inverse channel as
\begin{equation}
\begin{aligned}
\RM{\Gamma}_k &= \frac{\|\V{\alpha}_k\|_1}{N} \sum_{i=1}^{N} s_{\rv{l}_i}^{(k)}\M{O}_{\rv{l}_i}\\
&= \|\V{\alpha}_k\|_1 \sum_{l=1}^L s_l^{(k)} \widetilde{\rv{p}}_l^{(k)} \M{O}_l,
\end{aligned}
\end{equation}
where
$$
\widetilde{\rv{p}}_m^{(k)} = \frac{1}{N} \sum_{i=1}^{N} \mathbb{I}\{\rv{l}_i=m\}.
$$

The advantage of the Monte Carlo approach is that it may result in a much lower complexity for candidate circuit generation, compared to the exact implementation. To elaborate further, as a ``toy'' example, for a circuit consisting of two gates we have
\begin{equation}\label{two_gates_example}
\RM{\Gamma}_2\widetilde{\M{G}}_2\RM{\Gamma}_1\widetilde{\M{G}}_1= \frac{\|\V{\alpha}_1\|_1\|\V{\alpha}_2\|_1}{N}\sum_{i=1}^N s_{\rv{l}_{i,1}}^{(1)}s_{\rv{l}_{i,2}}^{(2)} \M{O}_{\rv{l}_{i,2}}\widetilde{\M{G}}_2 \M{O}_{\rv{l}_{i,1}}\widetilde{\M{G}}_1,
\end{equation}
where $\widetilde{\M{G}}_k = \M{C}_k\M{G}_k$, and $\rv{l}_{i,k}$ denotes the $i$-th sample drawn from the distribution $\V{p}_k$. This implies that in order to obtain a sample for the entire circuit, we may simply generate one sample for each gate, and concatenate them as shown in the right hand side of \eqref{two_gates_example}. Compared to the exact implementation, the Monte Carlo implementation can generate an arbitrary number of circuit samples $N$, at a relatively low computational cost of $O(NN_{\rm G})$.

The reduced complexity of the Monte Carlo implementation comes with a cost of inaccurate channel inversion, since $\M{G}_k$ is only an approximation of $\M{C}_k^{-1}$. Hence there would be a residual channel for each gate, which is given by
\begin{equation}
\widetilde{\RM{C}}_k = \RM{\Gamma}_k\M{C}_k.
\end{equation}
A natural question that arises is, whether the additional computational error caused by these residual channels would erode the error reduction capability of Monte Carlo-based \ac{qem}. In the rest of this treatise, we will discuss the impact of these residual channels on the accuracy vs. sampling overhead trade-off.

\section{Error Scaling Analysis of Monte Carlo-based \ac{qem}}\label{sec:analysis}
In this section, we discuss the error scaling behaviour of quantum circuits protected by Monte Carlo-based \ac{qem}, and contrast the results to that of circuits without \ac{qem} protection. In order to make a fair comparison, we consider the following assumptions.

\subsection{Assumptions}
\begin{assumption}[Bounded gate error rate]\label{asu:fidelity}
The error probability of each quantum gate is upper bounded by $\epsilon_{\rm u}$.
\end{assumption}

Since we consider Pauli channels in this treatise, the gate error probability corresponding to a quantum channel $\M{C}_k$ (under its \ac{ptm} representation) may be computed as
\begin{equation}
\epsilon(\M{C}_k) = 1-\frac{1}{4^n}\tr{\M{C}_k}.
\end{equation}

\begin{assumption}[Bounded observable]\label{asu:bounded}
The eigenvalues of the quantum observable $\Sop{M}_{\rm ob}$ are bounded in the interval $[-1,1]$.
\end{assumption}

Assumption \ref{asu:bounded} ensures the boundedness of the computation result $r$. In this treatise we assume that the upper and lower bounds are $1$ and $-1$, respectively, but they may be replaced with any other constant real numbers without affecting our analytical results. The assumption may also be rewritten as
\begin{equation}
\max_{\V{v}\in\mathbb{S}^n}\V{v}_{\rm ob}^{\rm T}\V{v}\V{v}^{\rm T} \V{v}_{\rm ob}\le 1,
\end{equation}
where $\mathbb{S}^n$ denotes the space of all density matrices over $n$ qubits. Furthermore, the assumption also implies that
\begin{equation}\label{normalization}
\|\V{v}_{\rm ob}\|_2 \le \sqrt{2^n}.
\end{equation}
This follows from the fact that $\|\V{v}_{\rm ob}\|_2 = \|\Sop{M}_{\rm ob}\|_{\rm F}$, and that
$$
\begin{aligned}
\|\Sop{M}_{\rm ob}\|_{\rm F} &= \sum_{i=1}^{2^n} \lambda_i(\Sop{M}_{\rm ob}) \\
&\le \sqrt{2^n},
\end{aligned}
$$
where $\lambda_i(\cdot)$ denotes the $i$-th largest eigenvalue of its argument.
\begin{assumption}[Zero bias term]\label{asu:bias}
We assume that
\begin{equation}
\tr{\Sop{M}_{\rm ob}} = \sqrt{2^n}[\V{v}_{\rm ob}]_1 = 0.
\end{equation}
\end{assumption}
Note that $[\V{v}_{\rm ob}]_1$ is the coefficient of the identity operator, which serves as a bias term in the computation result being constant with respect to the quantum state. Thus this assumption does not restrict the generality of our results.

\subsection{Benchmark: Error Scaling in the Absence of \ac{qem}}
In this subsections, we characterize the error scaling of quantum circuits that are not protected by \ac{qem}. The results will serve as important benchmarks in the following discussions. Let us start with a bound of the dynamic range of computational results, which will lead to a lower bound of the computational error.

\begin{proposition}\label{prop:noqem}
Assume that each qubit would be processed by at least $N_{\rm L}$ gates, and that for each of these gates, the probability of each type of Pauli error (i.e., X error, Y error or Z error) on each qubit is lower bounded by $\epsilon_{\rm l}$. The computational result $r$ exhibits the following convergence behaviour:
\begin{equation}\label{convergence_fp}
|r|\le \exp\left(-4\epsilon_{\rm l}N_{\rm L}\right).
\end{equation}
\begin{IEEEproof}
Please refer to Appendix \ref{sec:proof_noqem}.
\end{IEEEproof}
\end{proposition}

Proposition \ref{prop:noqem} implies that decoherence would force the computation result to be almost independent of the quantum observable $\V{v}_{\rm ob}$ in an asymptotic sense. Indeed, as indicated by \eqref{convergence_fp}, when $N_{\rm L}$ is large, $r$ is only determined by the first entry of $\V{v}_{\rm ob}$. Moreover, consider the case where $|\widetilde{r}| \ge 1-c$ holds for all $N_{\rm G}$, the computational error is lower bounded as
\begin{equation}
|r-\widetilde{r}| \ge 1-c - \exp\left(-4\epsilon_{\rm l}N_{\rm L}\right).
\end{equation}
From the Taylor expansion
$$
\exp\left(-4\epsilon_{\rm l}N_{\rm L}\right) = 1 - 4\epsilon_{\rm l}N_{\rm L} + \frac{(4\epsilon_{\rm l}N_{\rm L})^2}{2} - \cdots,
$$
we see that when $\epsilon_{\rm L}N_{\rm L}\ll 1$, the lower bound is approximately
\begin{equation}
|r-\widetilde{r}| \gtrapprox 4\epsilon_{\rm l}N_{\rm L} -c,
\end{equation}
which increases linearly with respect to $\epsilon_{\rm l}N_{\rm L}$.

We may also provide an upper bound for the computational error as follows.
\begin{proposition}\label{prop:noqem_ub}
The computational error can be upper bounded as
\begin{equation}\label{noqem_ub}
|r-\widetilde{r}| \le 2\epsilon_{\rm u}N_{\rm G}.
\end{equation}
\begin{IEEEproof}
Please refer to Appendix \ref{sec:proof_noqem_ub}.
\end{IEEEproof}
\end{proposition}

Combining Propositions \ref{prop:noqem} and \ref{prop:noqem_ub}, we see that the computational error grows linearly with $N_{\rm G}$, when the number of gates in each ``layer'' is constant (hence $N_{\rm L}$ is a constant multiple of $N_{\rm G}$. This is typically true for \acp{vqa}.

\subsection{The Statistics of the Residual Channels}
Before diving into details about the error scaling, in this subsection, we first investigate the characteristics of the residual channels of gates protected by Monte Carlo-based \ac{qem}.

According to the sampling overhead analysis in \cite{sof_analysis} based on the assumption of exact channel inversion, if we wish to execute the decoherence-free circuit $N_{\rm s}$ times, we should sample from the probabilistic mixture of candidate circuits for as many as
\begin{equation}
N = N_{\rm s}\|\V{\alpha}_k\|_1^2
\end{equation}
times, in order to keep the variance of the computational result unchanged by the channel inversion procedure. Here we consider the Monte Carlo-based channel inversion using the same number of samples, hence we have
\begin{equation}
\widetilde{\rv{p}}_m^{(k)} = \frac{1}{N_s\|\V{\alpha}_k\|_1^2} \sum_{i=1}^{N_s\|\V{\alpha}_k\|_1^2} \mathbb{I}\{\rv{l}_i=m\}.
\end{equation}
Of course, the Monte Carlo-based channel inversion has lower accuracy compared to the exact channel inversion, when they use the same number of samples. The accuracy could be improved by using additional samples, which will be discussed in more detail in Section \ref{ssec:ger}.

After the sampling procedure, $\V{\alpha}_k$ is approximated by $\widetilde{\RV{\alpha}}_k$ taking the following form
\begin{equation}
\begin{aligned}
\widetilde{\RV{\alpha}}_k &= \|\V{\alpha}_k\|_1\cdot \V{s}_k\odot \widetilde{\RV{p}}_k \\
&=\|\V{\alpha}_k\|_1\cdot \V{s}_k\odot(\V{p}_k + \RV{n}) \\
&=\V{\alpha}_k +  \|\V{\alpha}_k\|_1\cdot \V{s}_k\odot \RV{n},
\end{aligned}
\end{equation}
where $\RV{n}$ denotes the sampling error. In general, the approximated inverse channel may be expressed in terms of $\widetilde{\RV{\alpha}}_k$ as
\begin{equation}\label{gamma_k}
\RM{\Gamma}_k = \sum_{i=1}^L [\widetilde{\RV{\alpha}}_k]_i \M{O}_i,
\end{equation}
where $\{\M{O}_i\}_{i=1}^L$ is a set of operators forming a basis of the space where the imperfect gate $\M{C}_k\M{G}_k$ resides in. Interested readers may refer to Table 1 in \cite{sof_analysis} for an example of such operator sets. For the Pauli channels considered in this treatise, $\RM{\Gamma}$ has a simpler form. Specifically, using \eqref{pauli_as_diagonal}, the quasi-probability representation vector may be expressed as
$$
\V{\alpha}_k = \widetilde{\M{H}}^{-1}(1/\V{c}_k),
$$
where $\widetilde{\M{H}}$ is the Hadamard transform over $n$ qubits, and $\widetilde{\M{H}}^{-1}$ is the corresponding inverse transform given by $\widetilde{\M{H}}^{-1} = \frac{1}{4^n}\widetilde{\M{H}}$. The approximated inverse channel can now be expressed as
\begin{equation}
\RM{\Gamma}_k = \mathrm{diag}^{-1}\{\widetilde{\M{H}}\widetilde{\RV{\alpha}}_k\},
\end{equation}
and thus the residual channel takes the following form
\begin{equation}
\widetilde{\RM{C}}_k = \mathrm{diag}^{-1}\{\widetilde{\M{H}}\widetilde{\RV{\alpha}}_k\odot \V{c}_k\}.
\end{equation}
To simplify our further analysis, we introduce $\widetilde{\RV{c}}_k := \widetilde{\M{H}}\widetilde{\RV{\alpha}}_k\odot \V{c}_k$, where $\widetilde{\RV{c}}_k$ may be further expressed as
\begin{equation}
\begin{aligned}
\widetilde{\RV{c}}_k &= \V{1} + \|\V{\alpha}_k\|_1\cdot \widetilde{\M{H}}(\V{s}_k\odot \RV{n})\odot \V{c}_k\\
&=\V{1} + \|\V{\alpha}_k\|_1\cdot \V{c}_k\odot \widetilde{\RV{n}},
\end{aligned}
\end{equation}
and $\widetilde{\RV{n}}:=\widetilde{\M{H}}(\V{s}_k\odot \RV{n})$. Note that the vector $\widetilde{\RV{p}}_k$ is a multinomial distributed random vector, satisfying
\begin{equation}
\begin{aligned}
\mathbb{E}\{\widetilde{\RV{p}}_k\} &= \V{p}_k,~~\mathrm{Cov}\{\widetilde{\RV{p}}_k\} = \frac{1}{N_{\rm s}\|\V{\alpha}_k\|_1^2}\left(\M{P}_k-\V{p}_k\V{p}_k^{\rm T}\right),
\end{aligned}
\end{equation}
where $\M{P}_k=\diag{\V{p}_k}$. Therefore, the vector $\widetilde{\RV{n}}$ satisfies
\begin{equation}
\begin{aligned}
\mathbb{E}\{\widetilde{\RV{n}}\} &= \V{0},~~\mathrm{Cov}\{\widetilde{\RV{n}}\}=\widetilde{\M{H}}\mathrm{Cov}\{\widetilde{\RV{p}}_k\}\widetilde{\M{H}},
\end{aligned}
\end{equation}
since the sign vector $\V{s}_k$ does not have an effect on the covariance matrix. Thus we have the following results for $\widetilde{\RV{c}}_k$:
\begin{equation}\label{residual_distribution}
\begin{aligned}
\mathbb{E}\{\widetilde{\RV{c}}_k\} &= \V{1},\\
\mathrm{Cov}\{\widetilde{\RV{c}}_k\} &= \|\V{\alpha_k}\|_1^2 \cdot \mathrm{Cov}\{\V{c}_k\odot \widetilde{\RV{n}}\} \\
&=\frac{1}{N_{\rm s}} \widetilde{\M{H}}\left(\M{P}_k-\V{p}_k\V{p}_k^{\rm T}\right)\widetilde{\M{H}}\odot \V{c}_k\V{c}_k^{\rm T}.
\end{aligned}
\end{equation}
For simplicity of further derivation, we use the notation of $\M{\Xi}_k:=\mathrm{Cov}\{\widetilde{\RV{c}}_k\}$.

\subsection{Error Scaling in the Presence of Monte Carlo-based \ac{qem}}\label{ssec:qem}
In this subsection, we investigate the scaling law of computational error when the quantum circuit is protected by Monte Carlo-based \ac{qem}, based on the above discussions concerning the residual channels in the previous subsection.

We note that for \ac{qem}-protected circuits, the computational result is a random variable due to the randomness in the sampling procedure, given by
\begin{equation}
\rv{r} = \V{v}_{\rm ob}^{\rm T}\RV{v}_{N_{\rm G}},
\end{equation}
where
\begin{equation}\label{recursive_vk_random}
\RV{v}_k = \RM{R}_k \V{v}_0 =\widetilde{\RM{C}}_k\M{G}_k\RV{v}_{k-1}.
\end{equation}

After defining these quantities, we may obtain the following bound on the \ac{rmse} of the computational result $\rv{r}$.
\begin{proposition}[\color{sotonMarineBlue}Square-root Increase of \ac{qem} Inaccuracy]\label{prop:qem_perfectCSI}
For a quantum circuit consisting of $N_{\rm G}$ gates which is protected by \ac{qem}, the \ac{rmse} of the computational result is upper bounded by
\begin{equation}\label{qem_rmse_bound1}
\begin{aligned}
\sqrt{\mathbb{E}\{(\rv{r}-\widetilde{r})^2\}} &\le 2^{n/2}\sqrt{\exp(2N_{\rm G}N_{\rm s}^{-1})-1}.
\end{aligned}
\end{equation}
This result is not restricted to Pauli channels. In fact, it applies to all completely positive trace-preserving channels, when the basis operators $\{\M{O}_i\}_{i=1}^L$ are all completely positive trace-nonincreasing operators.
\begin{IEEEproof}
Please refer to Appendix \ref{sec:proof_qem_perfectCSI} for the proof of this proposition, as well as additional discussions on Pauli channels as a special case.
\end{IEEEproof}
\end{proposition}

Note that by applying the Taylor expansion to $\exp(2N_{\rm G}N_{\rm s})$, we have
$$
\exp(2N_{\rm G}N_{\rm s})-1 = \frac{2}{N_{\rm s}}N_{\rm G} + \frac{1}{2}\left(\frac{2}{N_{\rm s}}N_{\rm G}\right)^2 + \cdots,
$$
which is approximately $2N_{\rm G}N_{\rm s}^{-1}$, when $N_{\rm G}\ll N_{\rm s}$. This means that when the \ac{rmse} is far less than $1$, its scaling law is given by $O(\sqrt{N_{\rm G}}/\sqrt{N_{\rm s}})$. This is particularly useful, since in typical applications (e.g., variational quantum algorithms), having an \ac{rmse} close to 1 would be excessive.

In Proposition \ref{prop:qem_perfectCSI}, the dependence of the \ac{rmse} on the error probability of quantum gates is not demonstrated. According to \eqref{trace_ub_exact} of the Appendix, for Pauli channels, this dependence mainly relies on the term $\|\M{\Xi}_k\|_{\max}$. Next we expound a little further on this issue based on Assumption \ref{asu:fidelity}.

\begin{proposition}[Improved Bound for Pauli Channels]\label{prop:qem_ger}
Under Assumption \ref{asu:fidelity}, we have the following refined upper-bound for the \ac{rmse} of the computational result under Pauli channels:
\begin{equation}\label{qem_rmse_bound2}
\begin{aligned}
\sqrt{\mathbb{E}\{(\rv{r}-\widetilde{r})^2\}} &\le 2^{n/2}\sqrt{\exp\left(\tilde{\epsilon} N_{\rm G}N_{\rm s}^{-1}\right)-1} \\
&\approx 2^{n/2}\sqrt{\exp\left(10\epsilon_{\rm u} N_{\rm G}N_{\rm s}^{-1}\right)-1}
\end{aligned}
\end{equation}
where $\tilde{\epsilon}$ is given by
\begin{equation}\label{bound2_1}
\tilde{\epsilon} := \frac{5}{2}\sigma_{\rm u} + \frac{1}{4}\sigma_{\rm u}^2,
\end{equation}
and
\begin{equation}\label{bound2_2}
\sigma_{\rm u} := 4\epsilon_{\rm u}\cdot \frac{1-\epsilon_{\rm u}}{(1-2\epsilon_{\rm u})^2}.
\end{equation}
The approximation is valid when $\epsilon_{\rm u}\ll 1$.
\begin{IEEEproof}
Please refer to Appendix \ref{sec:proof_qem_ger}.
\end{IEEEproof}
\end{proposition}

Verification of the approximation in \eqref{qem_rmse_bound2} is straightforward: one may simply substitute \eqref{bound2_1} and \eqref{bound2_2} into \eqref{qem_rmse_bound2}. This proposition implies that, when $\epsilon_{\rm u}N_{\rm G} \ll N_{\rm s}$, the \ac{rmse} is on the order of $O(\sqrt{\epsilon_{\rm u}N_{\rm G}}/\sqrt{N_{\rm s}})$.

Engendered by our specific proof technique, the factor $2^{n/2}$ in \eqref{qem_rmse_bound1} and \eqref{qem_rmse_bound2} seems to be an artifact. According to the numerical results which will be presented in Section \ref{sec:numerical}, we conjecture this factor is essentially unnecessary, implying that
\begin{equation}\label{qem_conjecture}
\sqrt{\mathbb{E}\{(\rv{r}-\widetilde{r})^2\}} \le \sqrt{\exp\left(\epsilon N_{\rm G}N_{\rm s}^{-1}\right)-1}.
\end{equation}

Regretfully, it seems to be technically challenging to remove this factor from the bounds. Further investigations into this issue will be left for our future research.

\section{Discussions}\label{sec:discussion}

\subsection{Intuitions about the Error Scaling with the Circuit Size}
As indicated by the results in Section \ref{sec:analysis}, with respect to $N_{\rm G}$, we observe an $O(\sqrt{N_{\rm G}})$ scaling of the computational error of circuits protected by Monte Carlo-based \ac{qem}, when the number of samples is the same as that of \ac{qem} based on exact channel inversion. By contrast, when \ac{qem} is not applied, the computational error scales as $O(N_{\rm G})$, as discussed in Section \ref{sec:analysis}. Thus we may conclude that, although there are residual channels due to the inexact channel inversion, Monte Carlo-based \ac{qem} can still slow down the accumulation of computational error.

\begin{figure}[t]
\begin{minipage}{\columnwidth}
\centering
\begin{overpic}[width=.9\textwidth]{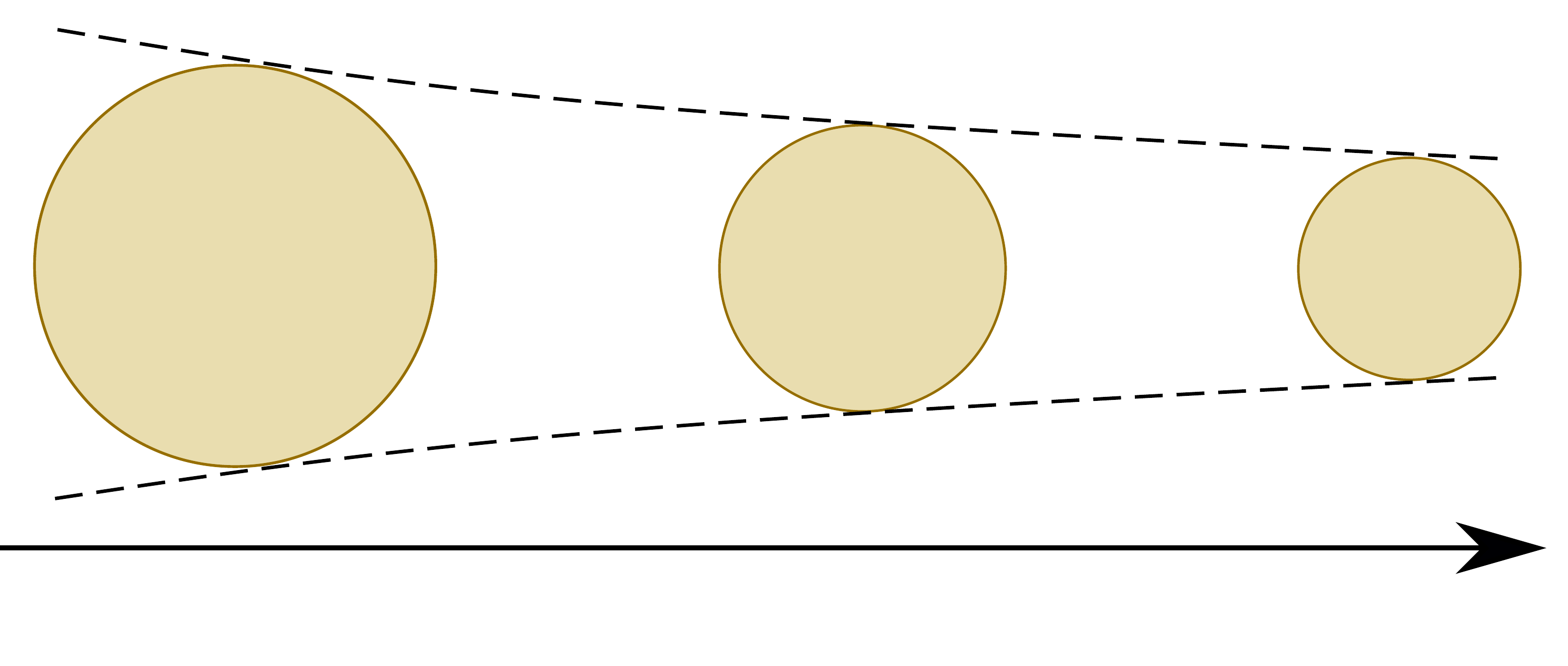}
\put(71,2.5){\footnotesize As $N_{\rm G}$ increases}
\end{overpic} \\
{\footnotesize (a) Gates without \ac{qem} protection}
\end{minipage}
\vspace{3mm}
\\
\vspace{3mm}
\begin{minipage}{\columnwidth}
\centering
\begin{overpic}[width=.9\textwidth]{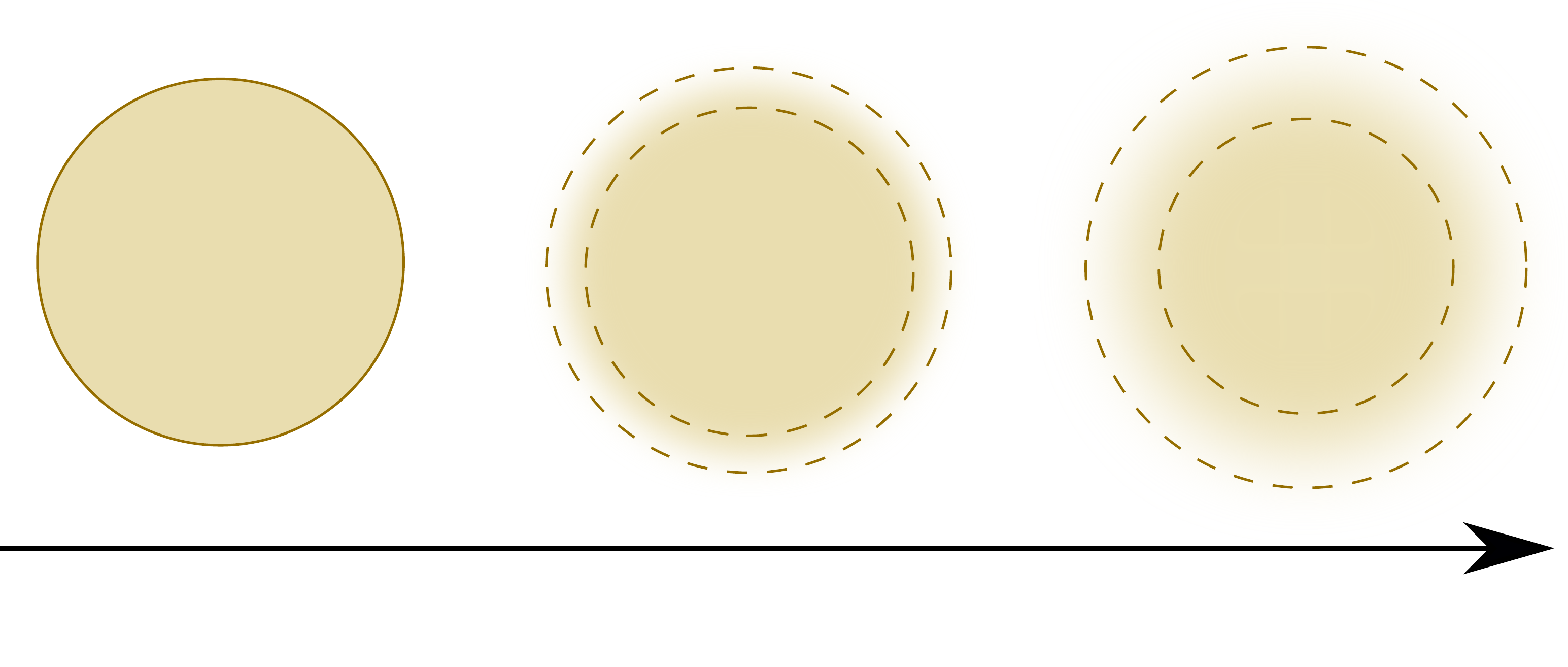}
\put(71,2.5){\footnotesize As $N_{\rm G}$ increases}
\end{overpic}\\
{\footnotesize (b) Gates protected by \ac{qem}}
\end{minipage}
\caption{Schematic illustration of the Bloch sphere undergoing a sequence of $N_{\rm G}$ imperfect single-qubit gates. The Bloch sphere shrinks when \ac{qem} is not applied, whereas it becomes ``blurred'' when the Monto Carlo-based \ac{qem} is applied.}
\label{fig:qem_vs_noqem}
\end{figure}

Revisiting the low-complexity example of a single-qubit circuit, we may understand these error scaling behaviours more intuitively. Specifically, the entire space of all legitimate single-qubit quantum states can be described by the celebrated Bloch sphere \cite[Sec. 1.2]{ncbook}. As demonstrated in Fig. \ref{fig:qem_vs_noqem}, the Bloch sphere would shrink as $N_{\rm G}$ increases when no \ac{qem} is applied, since the completely positive trace-preserving quantum channels are contractive transformations. This is in stark contrast with the case where Monte Carlo-based \ac{qem} is applied, when the Bloch sphere becomes ``blurred'' as $N_{\rm G}$ increases, since it is not determined whether the sphere will expand or shrink after each gate. Consequently, the sphere may expand after one gate and then shrink after another, hence the corresponding computational errors would cancel each other to a certain extent.

In light of the aforementioned intuition, we may interpret the error scaling of Monte Carlo-based \ac{qem} in following informal way. Assume that every gate $k$ would transform the Bloch sphere in a way that its radius becomes $(1+\rv{\lambda}_k)$ times that of its original value, where $\rv{\lambda}_k$ is a zero-mean random variable with variance $\sigma_k^2$. If additionally all $\rv{\lambda}_k$ values are mutually independent, we can see that
$$
\frac{1}{N_{\rm G}}\sum_{k=1}^{N_{\rm G}} \ln(1+\rv{\lambda}_k) \sim \mathcal{N}\left(-\frac{1}{2}\sigma^2,\sigma^2\right)
$$
holds asymptotically as $N_{\rm G}\rightarrow \infty$ by applying the central limit theorem, where we have:
$$
\sigma^2 = \frac{1}{N_{\rm G}}\sum_{k=1}^{N_{\rm G}} \sigma_k^2.
$$
Hence the radius of the Bloch sphere after $k$ gates, denoted by $\rv{a}_k$, tends to be a log-normally distributed random variable characterized by
$$
\begin{aligned}
\mathbb{E}\{\rv{a}_k\}&=1, \\
\mathrm{Var}\{\rv{a}_k\}&=\exp\left(N_{\rm G}\sigma^2\right)-1.
\end{aligned}
$$
Therefore, the standard deviation of the Bloch sphere's radius tends to be $\sqrt{\exp\left(N_{\rm G}\sigma^2\right)-1}$, which is on the order of $O\left(\sqrt{N_{\rm G}\sigma^2}\right)$ when $N_{\rm G}\sigma^2\ll 1$. This agrees with our formal analytical results.

\begin{figure}[t]
\centering
\begin{overpic}[width=0.45\textwidth]{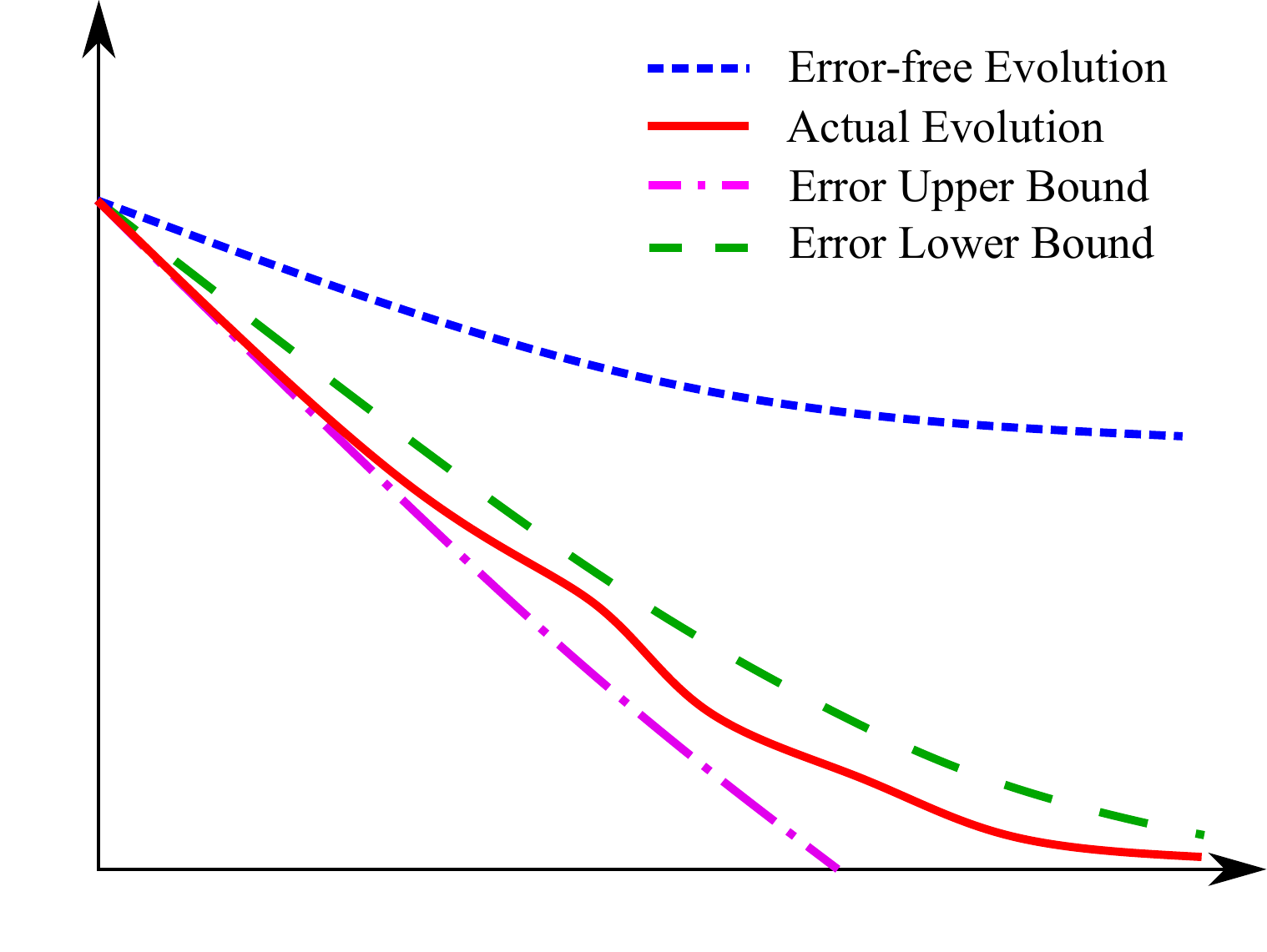}
\put(90,1.5){$N_{\rm G}$}
\put(3,70){$r$}
\put(4,4){$0$}
\end{overpic}
\caption{Demonstration of the evolution of computational result $r$ in the absence of \ac{qem}, as a function of $N_{\rm G}$. The upper and lower bound of computational error correspond to the results in Proposition \ref{prop:noqem_ub} and Proposition \ref{prop:noqem}, respectively.}
\label{fig:noqem_scaling}
\end{figure}

The linear error scaling experienced in the case where no \ac{qem} is applied may be interpreted by considering the graphical illustration in Fig. \ref{fig:noqem_scaling}. Since the computational result $r$ converges exponentially fast to zero as indicated by Proposition \ref{prop:noqem}, it deviates from $\widetilde{r}$ linearly when $N_{\rm G}$ is relatively small, which may be viewed as a lower bound of the computational error. Additionally, the actual evolution of $r$ is also bounded by the tangent line of it at $N_{\rm G}=0$, which gives rise to the error upper-bound in Proposition \ref{prop:noqem_ub}.

\subsection{The Accuracy vs. Sampling Overhead Trade-off}\label{ssec:ger}
If we denote by $\epsilon$ the average error probability of each gate, from the discussion in Section \ref{sec:analysis} we see that the computational error roughly scales as $\Theta(\epsilon)$ when \ac{qem} is not applied, whereas it scales as $O(\sqrt{\epsilon})$ when Monte Carlo-based \ac{qem} is applied, according to Section \ref{ssec:qem}. This may be understood by considering the variance of the samples, which is proportional to $\epsilon$. Hence the \ac{rmse} is proportional to $\sqrt{\epsilon}$.

Since $\epsilon$ is typically far less than $1$, it seems that Monte Carlo-based \ac{qem} has a less preferable performance. Nevertheless, it is noteworthy that the error scaling in the \ac{qem}-protected case is actually $O(\sqrt{\epsilon N_{\rm s}^{-1}})$, where the number $N_{\rm s}$ of effective circuit executions  is a configurable parameter. The $O(\sqrt{N_{\rm s}^{-1}})$ dependency on $N_{\rm s}$ originates from the fact that the sampling variance scales as $O(N_{\rm s}^{-1})$. Therefore, our results should not be viewed as indicating the superiority of non-\ac{qem}-based solutions. Rather, they should be viewed as a suggestion on the specific selection of $N_{\rm s}$, in the sense that it should be on the order of $\epsilon$ to ensure the error scaling is as beneficial as that of the family of non-\ac{qem}-based solutions.

Similarly, by increasing $N_{\rm s}$ as a function of $N_{\rm G}$, one could also improve the error scaling of Monte Carlo-based \ac{qem} with the circuit size. Indeed, since the error of Monte Carlo-based \ac{qem} scales as $O(\sqrt{N_{\rm G}N_{\rm s}^{-1}})$, we can choose an $N_{\rm s}$ that is proportional to $N_{\rm G}$ in order to attain a constant error with respect to $N_{\rm G}$. Note that the exact channel inversion also has a constant error with respect to $N_{\rm G}$ in the asymptotic limit of $N_{\rm G}\rightarrow \infty$. Therefore, using Monte Carlo-based \ac{qem}, we could use $N_{\rm G}$ times the number of samples to attain the same error scaling as that of \ac{qem} based on exact channel inversion. In practical scenarios, however, this may be an excessive sampling overhead. Fortunately, even if we use the same number of samples as that of the exact channel inversion, Monte Carlo-based \ac{qem} still exhibits a quadratic error scaling improvement compared to the no-\ac{qem}-based case.

\subsection{The Intrinsic Uncertainty of the Computational Results}
In the previous discussions, we followed the definition of computational results in \eqref{computational_result}. But even if the gates are decoherence-free, the intrinsic uncertainty of quantum states may bring some randomness to the computational result. To be specific, for a quantum state $\rho$, the variance of a quantum observable $\Sop{O}$ may be computed as follows \cite{vqe_theory}:
\begin{equation}
\mathrm{Var}_{\rho} \{\Sop{O}\} = \tr{\Sop{O}^2\rho} - (\tr{\Sop{O}\rho})^2,
\end{equation}
which quantifies the intrinsic uncertainty of the state $\rho$ under the observable $\mathcal{O}$. If the quantum circuit is executed $N_{\rm s}$ times, the variance is then given by $N_{\rm s}^{-1} \mathrm{Var}_{\rho} \{\Sop{O}\}$, and hence the \ac{mse} may be expressed as
\begin{equation}
\mathrm{MSE} = (r-\widetilde{r})^2+\frac{1}{N_{\rm s}}\cdot \mathrm{Var}_{\rho} \{\Sop{O}\}.
\end{equation}

We first consider the case where \ac{qem} is not applied. Since in \acp{vqa}, the observable $\Sop{M}_{\rm ob}$ is typically implemented using a Pauli operator decomposition, its variance may also be decomposed as
\begin{equation}
\mathrm{Var}_{\rho_{N_{\rm G}}}\{\Sop{M}_{\rm ob}\} = \sum_{i=1}^{4^n} \frac{1}{2^n}[\V{v}_{\rm ob}]_i^2 \mathrm{Var}_{\rho_{N_{\rm G}}}\{\Sop{S}_i\}.
\end{equation}
For each Pauli operator, we have
\begin{equation}
\begin{aligned}
\mathrm{Var}_{\rho_{N_{\rm G}}}\{\Sop{S}_i\} &=\tr{\Sop{S}_i^2\rho_{N_{\rm G}}} - (\tr{\Sop{S}_i\rho_{N_{\rm G}}})^2 \\
&= 1 -  (\tr{\Sop{S}_i\rho_{N_{\rm G}}})^2.
\end{aligned}
\end{equation}
Hence
\begin{equation}
\begin{aligned}
\mathrm{Var}_{\rho_{N_{\rm G}}}\{\Sop{M}_{\rm ob}\} &= \sum_{i=1}^{4^n} \frac{1}{2^n}[\V{v}_{\rm ob}]_i^2  (1 -  (\tr{\Sop{S}_i\rho_{N_{\rm G}}})^2)\\
&= \V{v}_{\rm ob}^{\rm T}\left(\frac{1}{2^n}\M{I}-\M{V}_{N_{\rm G}}^2\right)\V{v}_{\rm ob},
\end{aligned}
\end{equation}
where $\M{V}_{N_{\rm G}}=\diag{\V{v}_{N_{\rm G}}}$. Note that from \eqref{ptm_vector} we have $[\V{v}_{N_{\rm G}}]_i^2\le 2^{-n}$ for all $i$, hence it follows that
\begin{equation}
0\le \mathrm{Var}_{\rho_{N_{\rm G}}}\{\Sop{M}_{\rm ob}\} \le 1.
\end{equation}
Thus the \ac{mse} of the computational result is bounded by
\begin{equation}
(r-\widetilde{r})^2 \le \mathrm{MSE} \le (r-\widetilde{r})^2+\frac{1}{N_{\rm s}}.
\end{equation}

When circuits are protected by \ac{qem}, it has been shown that \cite{qem} if the number of effective executions is $N_{\rm s}$, the variance equals to that in the case where \ac{qem} is not applied. Thus the total error scales on the order of
$$
O\left(\sqrt{\frac{\epsilon N_{\rm G}}{N_{\rm s}}}\right) + O\left(\sqrt{\frac{1}{N_{\rm s}}}\right).
$$
This implies that, the effect of \ac{qem} may not be very significant when $\epsilon N_{\rm G}\ll 1$. But note that when $\rho_{N_{\rm G}}$ corresponds to one of the eigenstates of all Pauli operators $i$ having non-zero coefficient $[\V{v}_{\rm ob}]_i$, we have
$$
\mathrm{Var}_{\rho_{N_{\rm G}}}\{\Sop{S}_i\} = 1 - (\tr{\Sop{S}_i\rho_{N_{\rm G}}})^2 = 0,
$$
which follows from that fact that Pauli operators only have eigenvalues of $\pm 1$. Therefore, \ac{qem} would be more effective when the final state $\rho_{N_{\rm G}}$ is close to one of these eigenstates.

\section{Numerical Results}\label{sec:numerical}
In this section, we evaluate the analytical results presented in the previous sections via numerical examples. If not otherwise stated, the following parameters and assumptions will be used throughout the section.
\begin{itemize}
\item The number of effective circuit executions is $N_{\rm s}=5000$;
\item For Monte Carlo-based \ac{qem}, we use the same number of samples (i.e., actual circuit executions) as that of \ac{qem} based on exact channel inversion;
\item The quantum channels modelling the gate imperfections are single-qubit depolarising channels having gate error probability $10^{-3}$.
\end{itemize}

\subsection{Rotations Around the Bloch Sphere}\label{ssec:bloch}
\begin{figure}[t]
\subfloat[][Repeated Pauli X gates.]{
\begin{minipage}{.48\textwidth}
\centering
\includegraphics[width=.9\textwidth]{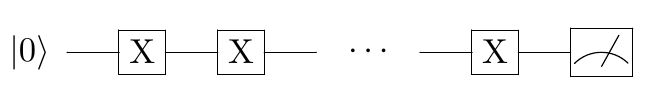}
\label{fig:repeated_X}
\end{minipage}
}
\\
\subfloat[][Repeated $\theta$-rotations around the X-axis.]{
\begin{minipage}{.48\textwidth}
\centering
\includegraphics[width=.9\textwidth]{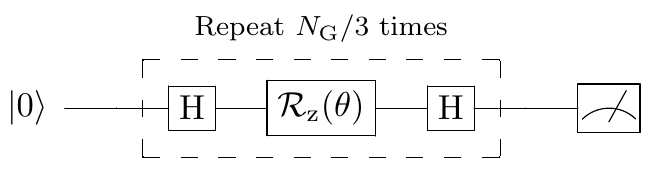}
\label{fig:repeated_rx}
\end{minipage}
}
\caption{Circuits implementing rotations around the X-axis of the Bloch sphere.}
\label{fig:circuit_rotation_bloch}
\end{figure}
We first consider the simplest scenario, where the quantum circuits are constituted of single-qubit gates, because these simple circuits allow us to clearly observe the error scaling described in the previous sections. In particular, we consider the circuits shown in Fig. \ref{fig:circuit_rotation_bloch}.  The quantum observable $\Sop{M}_{\rm ob}$ in this example is the Pauli Z operator $\Sop{Z}$ on the qubit, which satisfies
$$
\Sop{Z}\ket{0} = \ket{0},~\Sop{Z}\ket{1} = -\ket{1}.
$$
The corresponding \ac{ptm} representation is given by $\V{v}_{\rm ob} = [0~0~0~\sqrt{2}]^{\rm T}$.

\begin{figure}[t]
\centering
\subfloat[][\ac{rmse} versus $N_{\rm G}$.]{
\begin{minipage}{.46\textwidth}
\includegraphics[width=\textwidth]{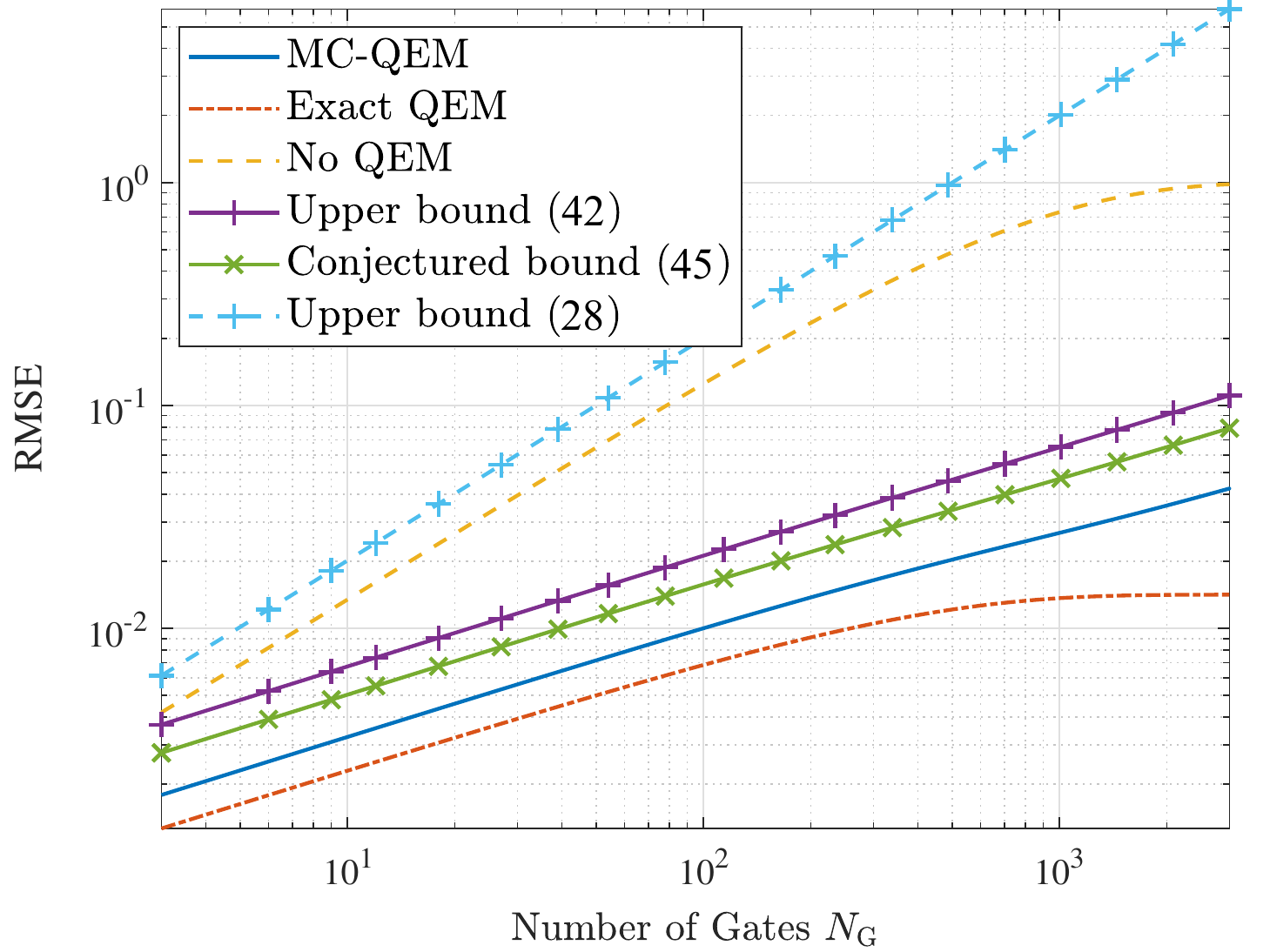}
\end{minipage}
\label{fig:bloch_rmse_X}
}
\\
\subfloat[][\ac{rmse} versus gate error probability ($N_{\rm G}=10$).]{
\begin{minipage}{.46\textwidth}
\includegraphics[width=\textwidth]{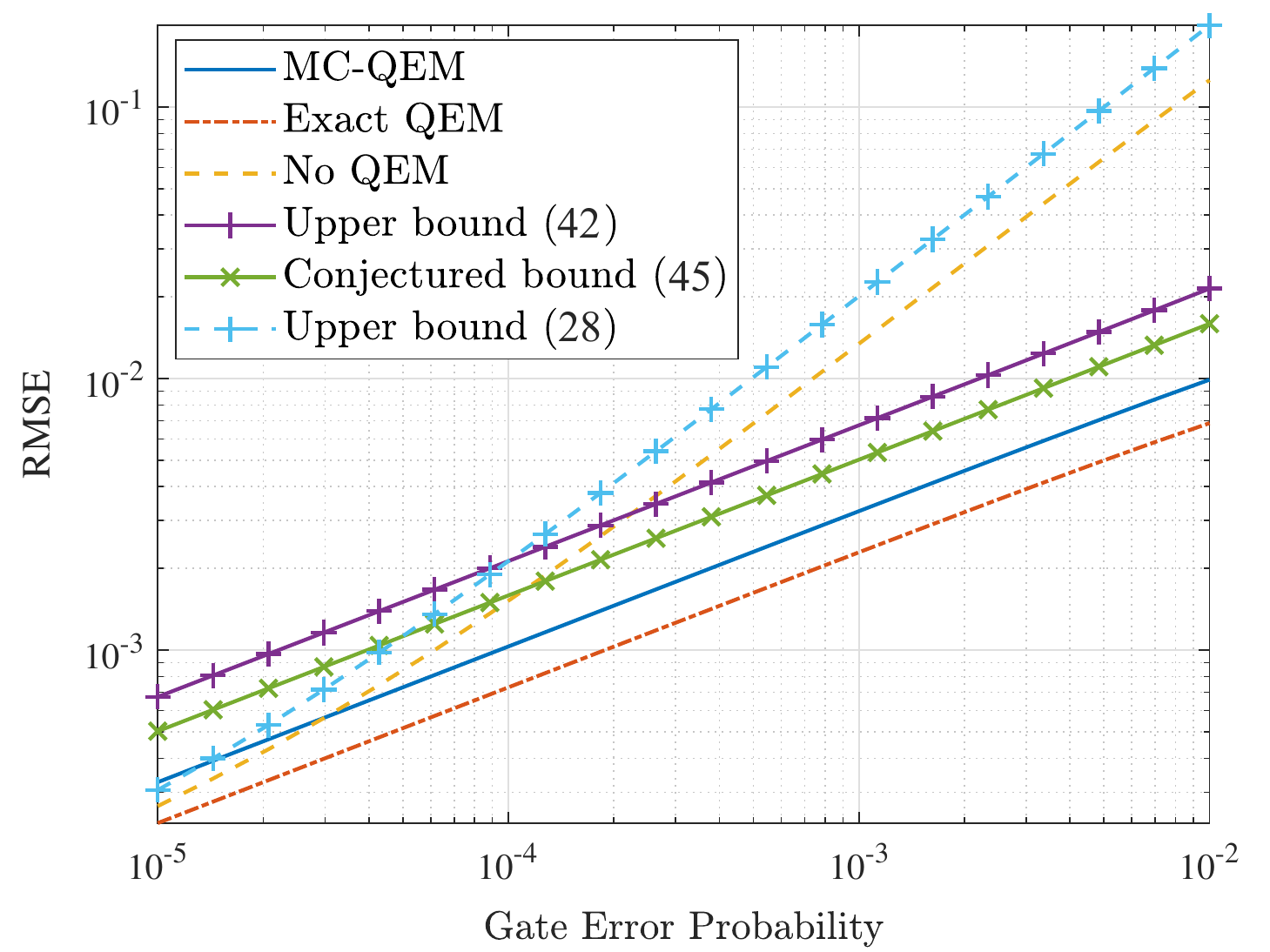}
\end{minipage}
\label{fig:bloch_rmse_X_epsilon}
}
\caption{The \ac{rmse} of the results computed by quantum circuits consisting of repeated Pauli X gates (as demonstrated in Fig. \ref{fig:repeated_X}).}
\label{fig:bloch_rmse}
\end{figure}

For the circuit consisting of repeated Pauli X gates shown in Fig. \ref{fig:repeated_X}, the \ac{rmse} of the computational results both with and without \ac{qem} protection is demonstrated in Fig. \ref{fig:bloch_rmse_X}, as a function of $N_{\rm G}$. As it can be observed from the figure, when $N_{\rm G}$ is relatively small, the \ac{rmse} of circuits operating without \ac{qem} protection grows linearly with $N_{\rm G}$, while the \ac{rmse} of circuits protected by Monte Carlo-based \ac{qem} scales as $O(\sqrt{N_{\rm G}})$.  The \ac{rmse} of \ac{qem} based on exact channel inversion scales as $O(\sqrt{N_{\rm G}})$ for small $N_{\rm G}$, but converges to a constant ($\approx \sqrt{N_{\rm s}^{-1}}$) when $N_{\rm G}$ is large.  Furthermore, when $N_{\rm G}$ is large, the \ac{rmse} of circuits operating without \ac{qem} protection converges to a constant. This agrees with Proposition \ref{prop:noqem}, which indicates that their computational results converge to zero regardless of the quantum observable.

The \ac{rmse} scalings with respect to the gate error probability $\epsilon$ are shown in Fig. \ref{fig:bloch_rmse_X_epsilon}, where we choose $N_{\rm G}=10$, while the number of effective circuit executions, namely $N_{\rm s}=5000$, does not vary as the gate error probability increases. It is noteworthy that when $\epsilon$ is small, the \ac{rmse} of circuits operating without \ac{qem} protection is lower than that of their counterparts protected by \ac{qem}. This phenomenon may be understood from our discussion in Section \ref{ssec:ger}, where we have indicated that the error scaling of \ac{qem}-protected circuits is $O(\sqrt{\epsilon N_{\rm s}^{-1}})$. Compared to the $O(\epsilon)$ scaling of non-\ac{qem}-protected circuits, the \ac{rmse} may be higher when $\epsilon$ is much smaller than $N_{\rm s}$. Interestingly, as seen from the figure, the square root scaling with respect to $\epsilon$ becomes preferable to the linear scaling when $\epsilon$ is relatively large.

\begin{figure}[t]
\centering
\begin{overpic}[width=0.46\textwidth]{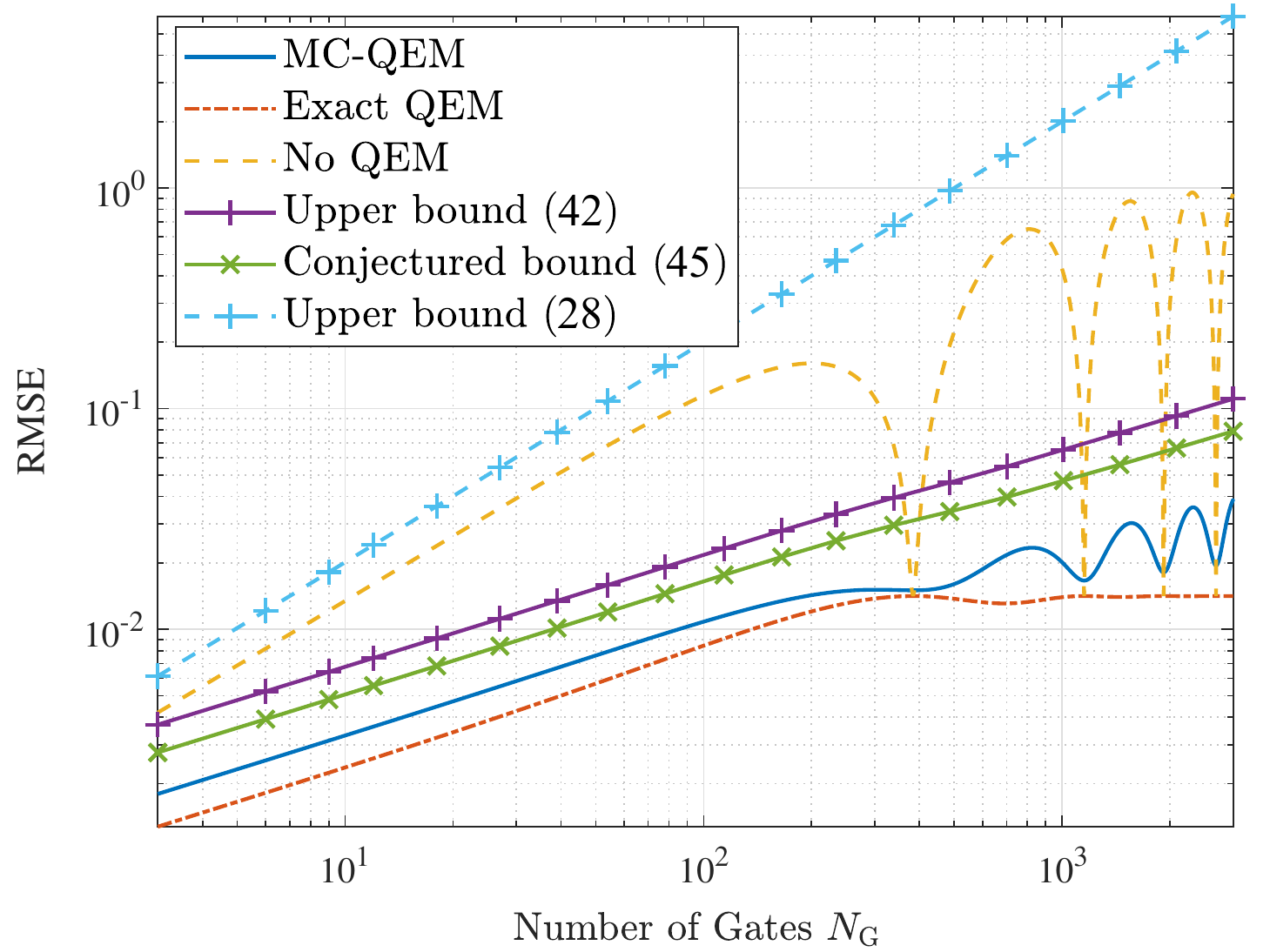}
\end{overpic}
\caption{The \ac{rmse} of the results computed by quantum circuits carrying out repeated rotations around the X-axis of the Bloch sphere (as shown in Fig. \ref{fig:repeated_rx}), as functions of $N_{\rm G}$, when $\theta=\pi/256$.}
\label{fig:bloch_rmse_rz}
\end{figure}

For the circuit comprising repeated rotations around the $X$-axis, as illustrated in Fig. \ref{fig:circuit_rotation_bloch}, we set $\theta = \pi/256$, and the results are plotted in Fig. \ref{fig:bloch_rmse_rz}. Observe that the envelope of the \ac{rmse} curves exhibit similar scaling behaviours as those in Fig. \ref{fig:bloch_rmse_X_epsilon}, but there are some oscillations. To understand the \ac{rmse} oscillations of circuits protected by Monte Carlo-based \ac{qem}, from \eqref{ak_recursion} we may express the covariance matrix of $\RV{v}_k$ as follows
\begin{equation}
\M{\Sigma}_k = (\V{1}\V{1}^{\rm T}+\M{\Xi}_k)\odot \M{G}_k\M{\Sigma}_{k-1}\M{G}_k^\dagger + \M{\Xi}_k\odot \V{\mu}_k\V{\mu}_k^{\rm T}.
\end{equation}
Note that the term $\M{\Xi}_k\otimes \V{\mu}_k\V{\mu}_k^{\rm T}$ varies with $k$ under the observable $\Sop{M}_{\rm ob}=\Sop{Z}$, and hence the \ac{rmse} is oscillatory.

\begin{figure}[t]
\centering
\begin{overpic}[width=0.46\textwidth]{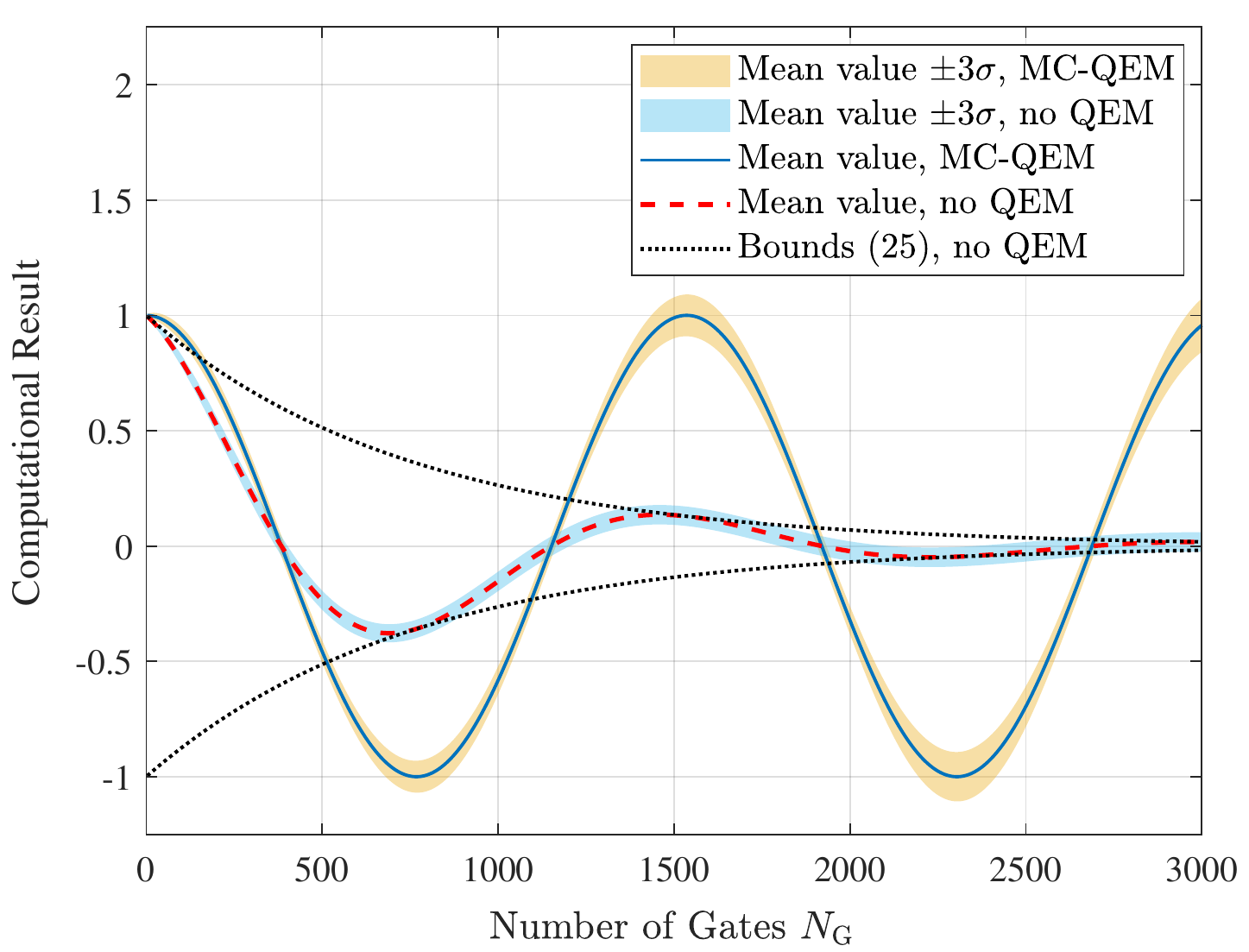}
\end{overpic}
\caption{The computational results of \ac{qem}-protected and non-\ac{qem}-protected circuits configured for carrying out repeated rotations ($\theta=\pi/256$) around the X-axis of the Bloch sphere (shown in Fig. \ref{fig:repeated_rx}), as functions of $N_{\rm G}$.}
\label{fig:bloch_mean}
\end{figure}

The \ac{rmse} oscillation of non-\ac{qem}-protected circuits may be better understood by investigating the evolution of the computational result $r$ as $N_{\rm G}$ increases, which is portrayed in Fig. \ref{fig:bloch_mean}. It can be seen that the mean values of the non-\ac{qem}-protected circuits fit nicely within the bounds given by Proposition \ref{prop:noqem}. Furthermore, the \ac{rmse} of \ac{qem}-protected circuits is mainly contributed by the variance of the computation results, while the \ac{rmse} of circuits not protected by \ac{qem} is mainly determined by the mean value, since in the latter case the bias is far larger than the standard deviation. As the dynamic range of the mean values is reduced, by coincidence, there are multiple intersections of the ground truth and the mean values, and thus the computational error of non-\ac{qem}-protected circuits oscillates as $N_{\rm G}$ increases.

\begin{figure}[t]
\centering
\subfloat[][Repeated Pauli X gates.]{
\begin{minipage}{.46\textwidth}
\includegraphics[width=\textwidth]{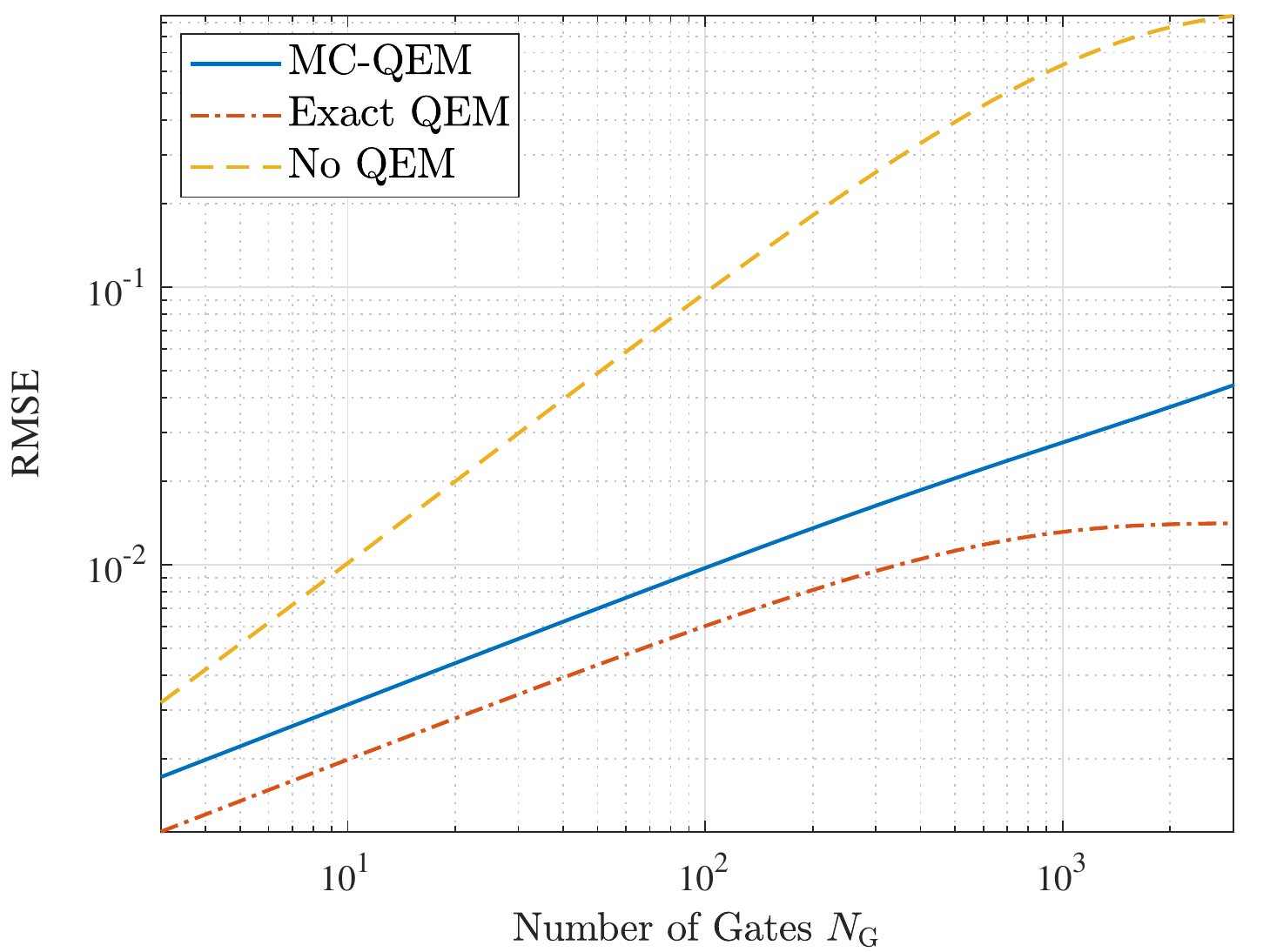}
\end{minipage}
\label{fig:bloch_rmse_X_damp}
}
\\
\subfloat[][Repeated $(\pi/256)$-rotations around the X-axis.]{
\begin{minipage}{.46\textwidth}
\includegraphics[width=\textwidth]{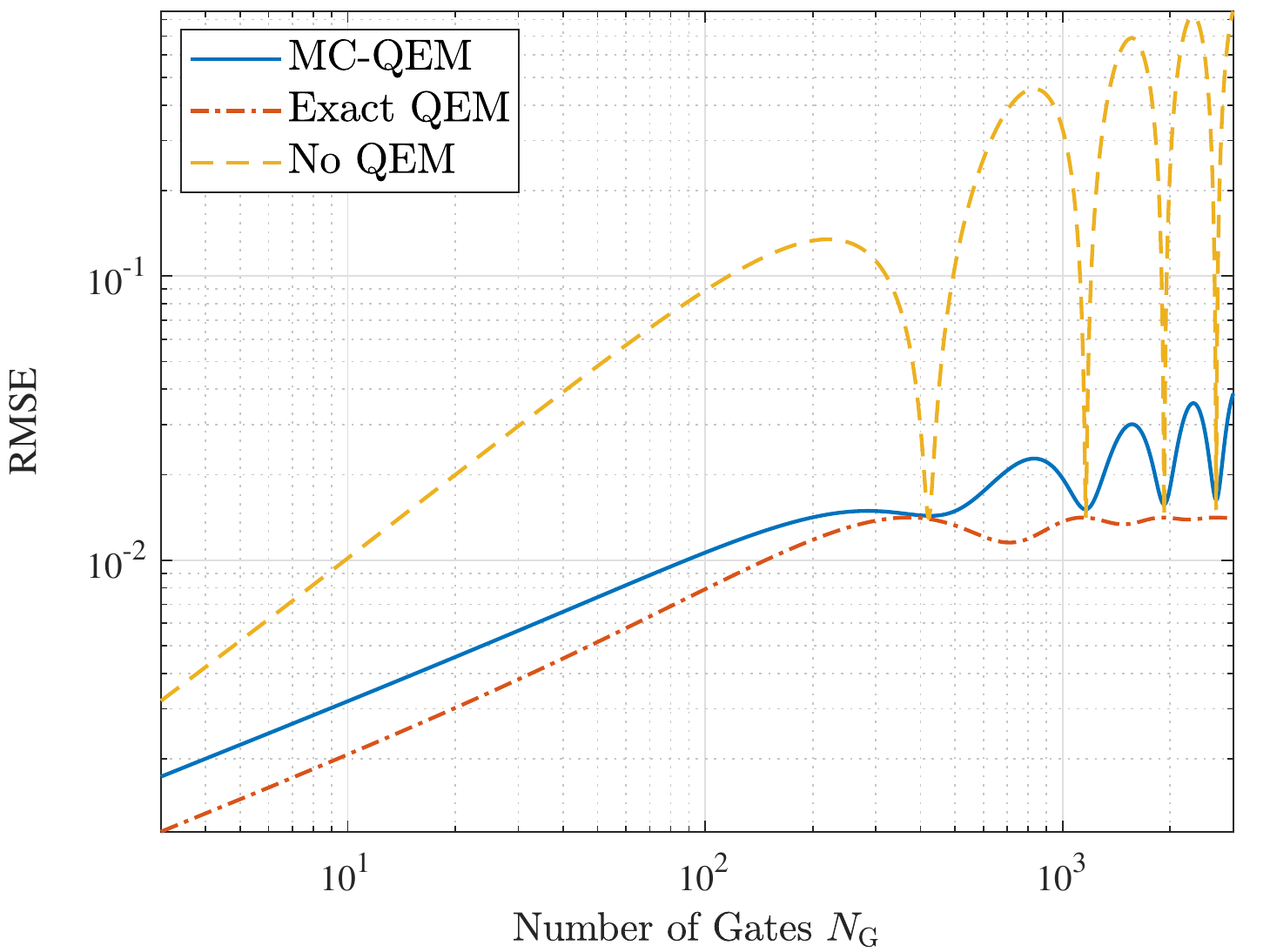}
\end{minipage}
\label{fig:bloch_rmse_rz_damp}
}
\caption{The \ac{rmse} of the results computed by quantum circuits demonstrated in Fig.~\ref{fig:circuit_rotation_bloch}, which are contaminated by amplitude damping channels.}
\end{figure}

Finally, we demonstrate that some non-Pauli channels may also exhibit the $O(\sqrt{N_{\rm G}})$ error scaling. In particular, we consider amplitude damping channels \cite{gst_intro} having the following \ac{ptm} representation
\begin{equation}
\M{C}_{\rm damp} = \left(
                                                                                                                                                      \begin{array}{cccc}
                                                                                                                                                        1 & 0 & 0 & 0 \\
                                                                                                                                                        0 & \sqrt{1-\gamma} & 0 & 0 \\
                                                                                                                                                        0 & 0 & \sqrt{1-\gamma} & 0 \\
                                                                                                                                                        \gamma & 0 & 0 & 1-\gamma \\
                                                                                                                                                      \end{array}
                                                                                                                                                    \right),
\end{equation}
where $\gamma$ is the amplitude damping probability. Here, we set the amplitude damping probability to $\gamma=1\times 10^{-3}$. The \ac{rmse} scalings with respect to the number of gates $N_{\rm G}$ are shown in Fig.~\ref{fig:bloch_rmse_X_damp} for the circuit comprising repeated Pauli X gates, and in Fig.~\ref{fig:bloch_rmse_rz_damp} for the circuit consisting of repeated $(\pi/256)$ rotations around the $X$-axis. We observe that the curves corresponding to \ac{qem} based on exact channel inversion and those corresponding to Monte Carlo-based \ac{qem} exhibit the $O(\sqrt{N_{\rm G}})$ scaling behavior, while the non-\ac{qem}-protected curves scale as $O(N_{\rm G})$, which is similar to the error scaling behaviour under Pauli channels as portrayed in Fig.~\ref{fig:bloch_rmse} and Fig.~\ref{fig:bloch_rmse_rz}.

\subsection{The Quantum Approximate Optimization Algorithm Aided Multi-User Detection}
In this subsection, we apply our analytical results to a practical variational quantum algorithm, the quantum approximate optimization algorithm \cite{qaoa}, which aims for solving combinatorial optimization problems assuming the following form
\begin{equation}\label{original_qaoa}
\max_{\V{z}\in\{-1,+1\}^n} ~~\sum_{k=1}^K w_k \prod_{i=1}^{n_k} z_{l_{k,i}},
\end{equation}
where $\V{z}=[z_1~\dotsc~z_n]^{\rm T}$, and $l_{k,i}\in\{1,2,\dotsc,n\}$. In the formulation of the quantum approximate optimization algorithm, the problem \eqref{original_qaoa} is transformed into the maximization of  $\bra{\psi}\Sop{H}\ket{\psi}$, where the quantum observable $\Sop{H}$ is given by
\begin{equation}\label{phase_hamiltonian}
\Sop{H} = \sum_{k=1}^K w_k \prod_{i=1}^{n_k} \Sop{Z}_{l_{k,i}}.
\end{equation}
The trial state $\ket{\psi}$ is prepared using a parametric circuit having an alternating structure, so that
$$
\ket{\psi} = e^{-i\beta_P\Sop{B}}e^{-i\gamma_P\Sop{H}} \cdots e^{-i\beta_1\Sop{B}}e^{-i\gamma_1\Sop{H}}\ket{+}^{\otimes n},
$$
where $P$ is the number of stages in the alternating circuit, and $\Sop{B}$ is the ``mixing Hamiltonian'' \cite{ansatz} given by $\Sop{B}=\sum_{i=1}^n \Sop{X}_i$. The parameters $\V{\beta}=[\beta_1~\dotsc~\beta_P]^{\rm T}$ and $\V{\gamma}=[\gamma_1~\dotsc~\gamma_P]^{\rm T}$ are typically obtained using via an optimization procedure implemented on classical computers \cite{vqe}. For the purpose of this treatise, here we do not optimize the parameters, but use the following (suboptimal) adiabatic configuration \cite{adiabatic} instead
$$
\gamma_k = kP^{-1},~~ \beta_k = 1- kP^{-1}.
$$

We consider the multiuser detection problem of wireless communications \cite{mud}. In particular, assuming that the modulation scheme is BPSK, in a spatial division multiple access system, the signal received at a base station equipped with $m$ antennas from $n$ single-antenna uplink transmitters may be expressed as
$$
\RV{y} = \M{H}\V{x} + \RV{\omega},
$$
where $\M{H}\in\mathbb{R}^{m\times n}$ denotes the channel, $\V{x}\in\{-1,+1\}^n$ represents the transmitted signal, and $\RV{\omega}\in\mathbb{R}^m$ is the noise. We assume here that the noise is i.i.d. Gaussian. Hence the maximum likelihood estimate of $\V{x}$ is given by
$$
\hat{\V{x}}_{\rm ML} = \mathop{\arg\max}_{\V{x}\in \{-1,+1\}^n} 2(\M{H}\RV{y})^{\rm T}\V{x} - \V{x}^{\rm T}\M{H}^{\rm T}\M{H}\V{x}.
$$
This may be further reformulated as the maximization of the quadratic form $\bra{\psi}\Sop{H}\ket{\psi}$, where
\begin{equation}\label{hamiltonian_mud}
\Sop{H} = \frac{1}{Z}\left(\sum_{i=1}^n [\M{H}^{\rm T}\RV{y}]_i\Sop{Z}_i -\sum_{i=1}^{n-1} \sum_{j=i+1}^n [\M{H}^{\rm T}\M{H}]_{i,j}\Sop{Z}_i\Sop{Z}_j\right),
\end{equation}
and $Z$ is a normalizing coefficient ensuring that the quantum observable $\Sop{H}$ satisfies our Assumption \ref{asu:bounded}.

\begin{figure}[t]
\centering
\begin{overpic}[width=0.46\textwidth]{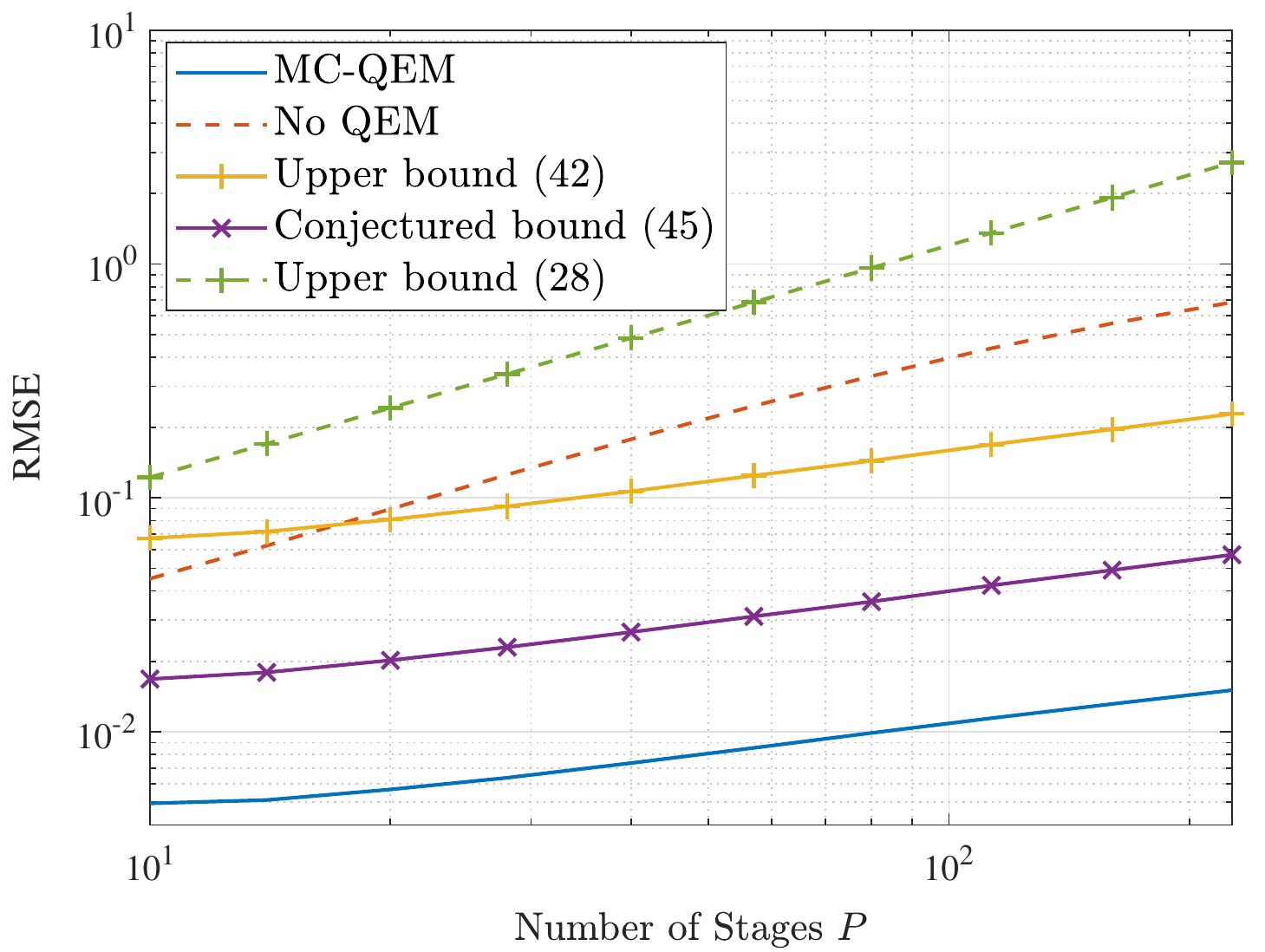}
\end{overpic}
\caption{The \ac{rmse} of the results computed by \ac{qem}-protected and non-\ac{qem}-protected circuits implementing the quantum approximate optimization algorithm based on \eqref{hamiltonian_mud}, as functions of the number of stages $P$.}
\label{fig:qaoa_rmse}
\end{figure}

In this illustrative example, we consider the case where $m=n=4$, and $[\RV{\omega}]_i\sim\Sop{N}(0,0.0631)$, $\forall i$, such that the signal-to-noise ratio is 12dB. We assume furthermore that the channels between each pair of antennas are uncorrelated non-dispersive Rayleigh channels, hence the entries of the channel $\M{H}$ are i.i.d. Gaussian variables with zero mean and variance $m^{-1}$ \cite{quantum_mud}. For the quantum circuits, we choose gate error probability $\epsilon=3\times 10^{-4}$. Under these assumptions, the \ac{rmse} scalings with respect to $P$ of non-\ac{qem}-protected circuits and that of circuits protected by Monte Carlo-based \ac{qem} are portrayed in Fig. \ref{fig:qaoa_rmse}. It can be observed that the non-\ac{qem}-protected circuits exhibit an $O(P)$ scaling, while the \ac{qem}-protected circuits exhibit an $O(\sqrt{P})$ scaling, as indicated by Propositions \ref{prop:noqem_ub} and \ref{prop:qem_ger}, respectively.

\begin{figure}[t]
\centering
\begin{overpic}[width=0.46\textwidth]{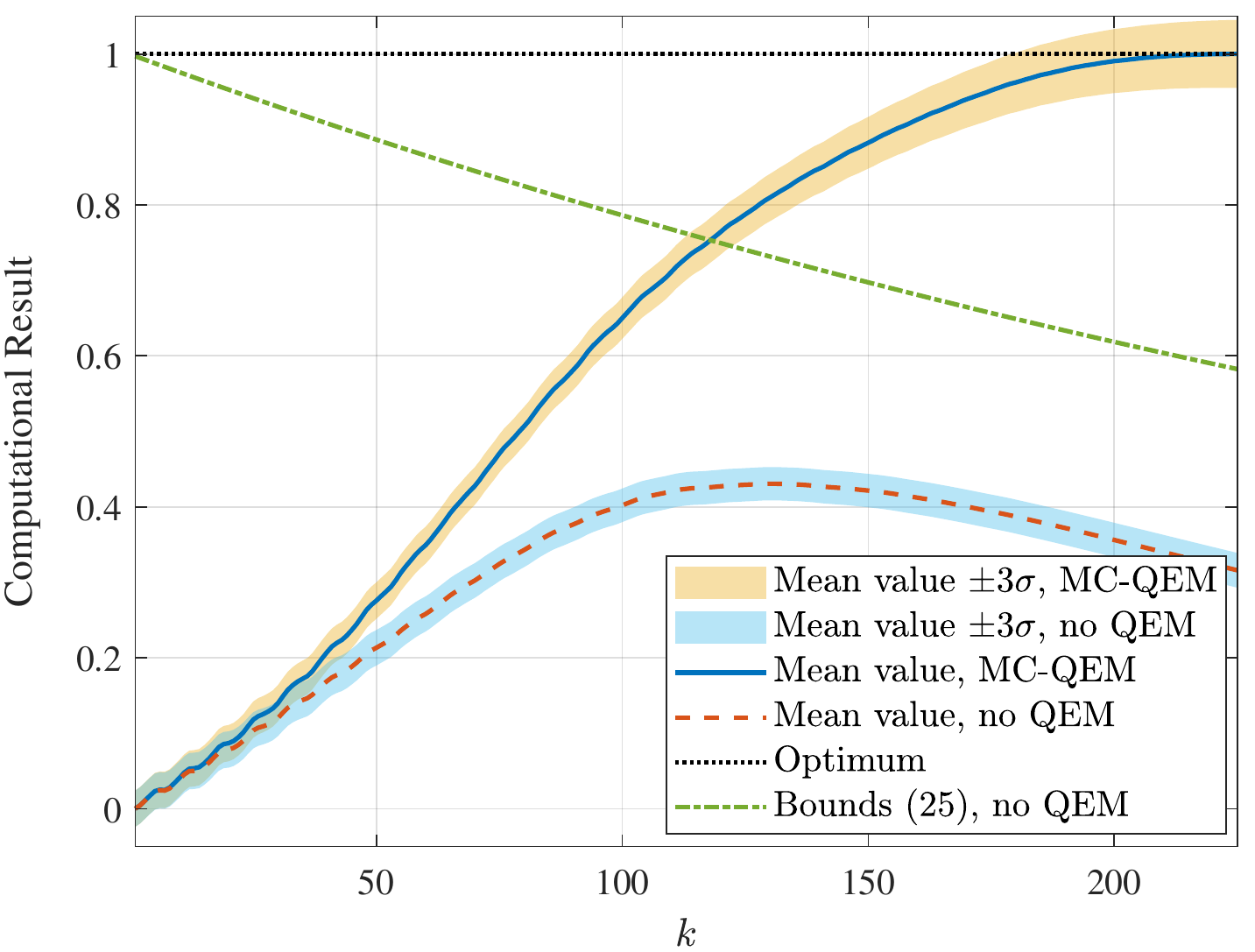}
\end{overpic}
\caption{The objective function values computed at the $k$-th stage of the quantum approximate optimization algorithm (for which $P=225$) implemented based on \eqref{hamiltonian_mud}.}
\label{fig:qaoa_mean_values}
\end{figure}

To illustrate the evolution of the computational results during the execution of circuits, we plot the objective function values (i.e., $\bra{\psi}\Sop{H}\ket{\psi}$) computed at each stage $k$ of the circuits in Fig. \ref{fig:qaoa_mean_values}, for the case where $P=225$. Note that the results computed by the \ac{qem}-protected circuits converge monotonically towards the optimum, for which the main source of error is the variance. By contrast, for the non-\ac{qem}-protected circuits, the results were on the right track for $k<100$, but soon they deviate from their \ac{qem}-protected counterparts, and start to converge to zero. In this example, the bound \eqref{convergence_fp} is not as tight as it was in Section \ref{ssec:bloch}, but it still indicates that the dynamic range of the results computed by non-\ac{qem}-protected circuits decays exponentially as $k$ increases.

\section{Conclusions}\label{sec:conclusions}
The trade-off between the computational overhead and the error scaling behaviour of both quantum circuits protected by Monte Carlo-based \ac{qem} and their non-\ac{qem}-protected counterparts was investigated. As for the non-\ac{qem}-protected circuits, we have shown that the dynamic range of the noisy computational results shrinks exponentially as the number of gates $N_{\rm G}$ increases, implying a linear error scaling with $N_{\rm G}$. By contrast, the error scales as the square root of $N_{\rm G}$ in the presence of Monte Carlo-based \ac{qem}, at the same computational cost as that of \ac{qem} based on exact channel inversion. Moreover, the error scaling of Monte Carlo-based \ac{qem} can be further improved with increased computational cost.

We have also demonstrated the analytical results both for low-complexity examples and for a more practical example of the quantum approximate optimization algorithm employed for multi-user detection in wireless communications. It may be an interesting future research direction to apply the results to other practical examples, or verify them using experimental approaches.

\section*{Acknowledgments}
Y. Xiong would like to acknowledge Daryus Chandra for helpful conversations regarding quantum channel estimation and quantum error correction codes.

\appendices
\section{Proof of Proposition \ref{prop:noqem}}\label{sec:proof_noqem}
\begin{IEEEproof}
First observe that the matrix representation of a perfect gate $\M{G}_i$ as well as that of a channel $\M{C}_i$ take the following block-diagonal form
\begin{equation}
\M{G}_i = \left(
            \begin{array}{cc}
              1 & \V{0}^{\rm T} \\
              \V{0} & \M{U}_i \\
            \end{array}
          \right),~~\M{C}_i=\left(
                              \begin{array}{cc}
                                1 & \V{0}^{\rm T} \\
                                \V{0} & \M{D}_i \\
                              \end{array}
                            \right),
\end{equation}
where $\M{U}_i$ is a unitary matrix, whereas $\M{D}_i$ is a diagonal matrix having diagonal entries taking values in the interval $[0,1]$. Since the matrix $\M{R}_{N_{\rm G}}$ is the product of several $\M{G}_i$ and $\M{C}_i$, it becomes clear that its largest singular values satisfies $\sigma_1(\M{R}_{N_{\rm G}})=1$, and its second largest singular value satisfies
\begin{equation}
\sigma_2(\M{R}_{N_{\rm G}})\le \prod_{i=1}^{N_{\rm G}} \|\M{D}_i\|_2.
\end{equation}
Furthermore, we have
\begin{equation}
\left|r-\frac{1}{2^n}\tr{\Sop{M}_{\rm ob}}\right|\le \sigma_2(\M{R}_{N_{\rm G}})
\end{equation}
due to the ``bounded observable'' Assumption \ref{asu:bounded}.

Note that the quantity $N_{\rm L}$ defined in this proposition is related to the depth of the circuit. To elaborate, if we say that ``a layer of gates'' is executed if each qubit has been processed by at least one gate, then the entire circuit consists of at least $N_{\rm L}$ layers. For each single-qubit channel $\M{C}$ in these layers, due to the assumption that each single-qubit Pauli error occurs at probability of at least $\epsilon_{\rm l}$, the following bound holds:
\begin{equation}
\begin{aligned}
\M{C} &= \M{I}-2\diag{[p_{\rm X}+p_{\rm Z}~p_{\rm Y}+p_{\rm Z}~p_{\rm X}+p_{\rm Y}]} \\
&\preceq (1-4\epsilon_{\rm l}),
\end{aligned}
\end{equation}
where $p_{\rm X}$, $p_{\rm Y}$ and $p_{\rm Z}$ are error probabilities corresponding to the Pauli-X, Y and Z errors, respectively. Thus we obtain
\begin{equation}
\begin{aligned}
\sigma_2(\M{R}_{N_{\rm G}})&\le (1-4\epsilon_{\rm l})^{N_{\rm L}}\\
&=\exp\{N_{\rm L}\ln(1-4\epsilon_{\rm l})\} \\
&\le \exp(-4\epsilon_{\rm l}N_{\rm L}).
\end{aligned}
\end{equation}
Hence the proof is completed.
\end{IEEEproof}

\section{Proof of Proposition \ref{prop:noqem_ub}}\label{sec:proof_noqem_ub}
\begin{IEEEproof}
In this proof, we will work under the operator-sum representation of quantum channels. Since we consider Pauli channels, the recursion \eqref{recursive_os} may be rewritten as
\begin{equation}\label{recursion_pauli}
\begin{aligned}
\rho_{k} &= \sum_{i=1}^{4^n} [\V{p}_k]_i \Sop{S}_i \Sop{G}_k\rho_{k-1}\Sop{G}_k^{\dagger} \Sop{S}_i \\
&=[\V{p}_k]_1 \Sop{G}_k\rho_{k-1}\Sop{G}_k^{\dagger} +  \sum_{i=2}^{4^n} [\V{p}_k]_i \Sop{S}_i \Sop{G}_k\rho_{k-1}\Sop{G}_k^{\dagger} \Sop{S}_i.
\end{aligned}
\end{equation}
Assumption \ref{asu:bounded} implies that $\|\Sop{M}_{\rm ob}\|_2\le 1$, meaning that
$$
\tr{\Sop{M}_{\rm ob}\rho} \le 1
$$
holds for any legitimate density matrix $\rho$. Note that terms such as $\Sop{S}_i \Sop{G}_k\rho_{k-1}\Sop{G}_k^{\dagger} \Sop{S}_i$ in \eqref{recursion_pauli} are indeed legitimate density matrices. Thus the computational result satisfies
\begin{equation}
\begin{aligned}
\left|\widetilde{r}-\tr{\Sop{M}_{\rm ob}\rho_{N_{\rm G}}}\right| &\le \left(|\widetilde{r}|+\|\Sop{M}_{\rm ob}\|_2\right)\left(1-\prod_{k=1}^{N_{\rm G}} [\V{p}_k]_1\right) \\
&\le 2\left(1-\prod_{k=1}^{N_{\rm G}} [\V{p}_k]_1\right).
\end{aligned}
\end{equation}
According to Assumption \ref{asu:fidelity}, for any $k$, the vector $\V{p}_k$ satisfies
\begin{equation}
\begin{aligned}
{[\V{p}_k]}_1 &\ge 1-\epsilon_{\rm u}, \\
\sum_{i=2}^{4^n} [\V{p}_k]_i &\le \epsilon_{\rm u}.
\end{aligned}
\end{equation}
Therefore, we have
\begin{equation}
\begin{aligned}
\left|\widetilde{r}-\tr{\Sop{M}_{\rm ob}\rho_{N_{\rm G}}}\right| &\le 2 (1-(1-\epsilon_{\rm u})^{N_{\rm G}}) \\
&\le 2\epsilon_{\rm u}N_{\rm G}.
\end{aligned}
\end{equation}
Hence the proof is completed.
\end{IEEEproof}

\section{Proof of Proposition \ref{prop:qem_perfectCSI}}\label{sec:proof_qem_perfectCSI}
\begin{IEEEproof}
We first expand the expression of MSE as follows
\begin{equation}\label{mse_qem}
\begin{aligned}
\mathbb{E}\{(\rv{r}-\widetilde{r})^2\}&=\mathbb{E}\left\{\left(\V{v}_{\rm ob}^{\rm T}\RV{v}_{N_{\rm G}}-\widetilde{r}\right)^2\right\} \\
&=\V{v}_{\rm ob}^{\rm T}\mathbb{E}\left\{\RV{v}_{N_{\rm G}}\RV{v}_{N_{\rm G}}^{\rm T}\right\}\V{v}_{\rm ob}+\widetilde{r}^2-2\widetilde{r}\V{v}_{\rm ob}^{\rm T}\mathbb{E}\{\RV{v}_{N_{\rm G}}\}.
\end{aligned}
\end{equation}
Hence the \ac{rmse} is given by
\begin{equation}
\sqrt{\mathbb{E}\{(\rv{r}-\widetilde{r})^2\}} = \sqrt{\V{v}_{\rm ob}^{\rm T}\M{A}_k\V{v}_{\rm ob}+\widetilde{r}^2-2\widetilde{r}\V{v}_{\rm ob}^{\rm T}\V{\mu}_k},
\end{equation}
where $\M{A}_k:=\mathbb{E}\{\RV{v}_k\RV{v}_k^{\rm T}\}$ and $\V{\mu}_k:=\mathbb{E}\{\RV{v}_k\}$.

Using \eqref{gamma_k} and \eqref{recursive_vk_random}, we have
\begin{equation}
\RV{v}_{k} = \sum_{i=1}^L[\widetilde{\RV{\alpha}}_k]_i\M{O}_i\M{C}_k\M{G}_k\RV{v}_{k-1}.
\end{equation}
This implies the following recursive relationships:
\begin{subequations}\label{recursion_A_nonpauli}
\begin{align}
\M{A}_k &= \sum_{i=1}^L\sum_{j=1}^L e_{ij}^{(k)}\M{O}_i\M{C}_k\M{G}_k\M{A}_{k-1}\M{G}_k^{\rm T}\M{C}_k^{\rm T}\M{O}_j^{\rm T}, \label{recursion_A_line1}\\
\V{\mu}_k &= \M{G}_k\V{\mu}_{k-1},
\end{align}
\end{subequations}
where $e_{ij}^{(k)}=[\M{E}_k]_{ij}:=\mathbb{E}\{[\widetilde{\RV{\alpha}}_k]_i[\widetilde{\RV{\alpha}}_k]_j\}$. The matrix $\M{E}_k$ may be expressed as
\begin{equation}
\begin{aligned}
\M{E}_k &= \mathbb{E}\{\widetilde{\RV{\alpha}}_k\widetilde{\RV{\alpha}}_k^{\rm T}\}\\
&=\V{\alpha}_k\V{\alpha}_k^{\rm T}+\|\V{\alpha}_k\|_1^2{\rm Cov}\{\widetilde{\RV{p}}_k\}\\
&=\V{\alpha}_k\V{\alpha}_k^{\rm T}+ \frac{1}{N_{\rm s}}\left(\M{P}_k-\V{p}_k\V{p}_k^{\rm T}\right).
\end{aligned}
\end{equation}
For the simplicity of further derivation, we denote $\widetilde{\M{E}}_k:=\frac{1}{N_{\rm s}}\left(\M{P}_k-\V{p}_k\V{p}_k^{\rm T}\right)$.

Let us now consider the case of $k=1$. In this case, $\V{v}_0$ of \eqref{ptm_result} is a deterministic vector, thus we have
\begin{equation}\label{initial_conditions}
\begin{aligned}
\M{A}_0 &= \V{v}_0\V{v}_0^{\rm T}, \\
\V{\mu}_0 &= \V{v}_0.
\end{aligned}
\end{equation}
Using the recursive relationship of $\V{\mu}_k=\M{G}_k\V{\mu}_{k-1}$, we now see that $\V{v}_{\rm ob}^{\rm T}\V{\mu}_{N_{\rm G}} = \widetilde{r}$. Hence we may simplify \eqref{mse_qem} as
\begin{equation}\label{covariance_substitution}
\begin{aligned}
\mathbb{E}\{(\rv{r}-\widetilde{r})^2\} &= \V{v}_{\rm ob}^{\rm T} \M{A}_{N_{\rm G}} \V{v}_{\rm ob} +\widetilde{r}^2-2\widetilde{r}\V{v}_{\rm ob}^{\rm T} \V{\mu}_{N_{\rm G}} \\
&=\V{v}_{\rm ob}^{\rm T} \M{A}_{N_{\rm G}} \V{v}_{\rm ob} -(\V{v}_{\rm ob}^{\rm T}\V{\mu}_{N_{\rm G}})^2 \\
&=\V{v}_{\rm ob}^{\rm T}\left(\M{A}_{N_{\rm G}}-\V{\mu}_{N_{\rm G}}\V{\mu}_{N_{\rm G}}^{\rm T}\right)\V{v}_{\rm ob}.
\end{aligned}
\end{equation}
Observe that the term $\M{A}_{N_{\rm G}}-\V{\mu}_{N_{\rm G}}\V{\mu}_{N_{\rm G}}^{\rm T}$ is in fact the covariance matrix of $\RV{v}_{N_{\rm G}}$, upon defining
\begin{equation}
\M{\Sigma}_k:= \M{A}_k-\V{\mu}_k\V{\mu}_k^{\rm T},
\end{equation}
and substituting into \eqref{covariance_substitution} we arrive at
\begin{equation}
\mathbb{E}\{(\rv{r}-\widetilde{r})^2\} = \V{v}_{\rm ob}^{\rm T}\M{\Sigma}_{N_{\rm G}}\V{v}_{\rm ob}.
\end{equation}
The covariance matrix can be further formulated as
\begin{equation}\label{sigma_k}
\begin{aligned}
\M{\Sigma}_k &= \M{A}_k - \V{\mu}_k\V{\mu}_k^{\rm T} \\
&=\M{A}_k - \widetilde{\M{R}}_k\V{v}_0\V{v}_0^{\rm T}\widetilde{\M{R}}_k^{\rm T}.
\end{aligned}
\end{equation}

It now suffices to compute $\M{A}_k$. Taking trace from both sides of \eqref{recursion_A_line1}, we have
\begin{equation}\label{trace_ub_exact_nonpauli}
\begin{aligned}
\tr{\M{A}_k}&=\sum_{i=1}^L\sum_{j=1}^L e_{ij}^{(k)}\tr{\M{O}_i\M{C}_k\M{G}_k\M{A}_{k-1}\M{G}_k^{\rm T}\M{C}_k^{\rm T}\M{O}_j^{\rm T}}.
\end{aligned}
\end{equation}
Next we consider the decomposition
\begin{equation}
e_{ij}^{(k)} = [\V{\alpha}_k]_i[\V{\alpha}_k]_j+\left[\widetilde{\M{E}}_k\right]_{ij}.
\end{equation}
Observe that the term $[\V{\alpha}_k]_i[\V{\alpha}_k]_j$ satisfies
\begin{equation}
\begin{aligned}
&\sum_{i=1}^L\sum_{j=1}^L[\V{\alpha}_k]_i[\V{\alpha}_k]_j\tr{\M{O}_i\M{C}_k\M{G}_k\M{A}_{k-1}\M{G}_k^{\rm T}\M{C}_k^{\rm T}\M{O}_j^{\rm T}}\\
&\hspace{3mm}=\tr{\M{A}_{k-1}},
\end{aligned}
\end{equation}
since $\M{C}_k^{-1}=\sum_{i=1}^L[\V{\alpha}_k]_i\M{O}_i$. Thus we have
\begin{equation}
\begin{aligned}
&\tr{\M{A}_k}-\tr{\M{A}_k}\\
&\hspace{3mm}=\sum_{i=1}^L\sum_{j=1}^L \left[\widetilde{\M{E}}_k\right]_{ij}\tr{\M{O}_i\M{C}_k\M{G}_k\M{A}_{k-1}\M{G}_k^{\rm T}\M{C}_k^{\rm T}\M{O}_j^{\rm T}}\\
&\hspace{3mm}\le \sum_{i=1}^L\sum_{j=1}^L \left[\widetilde{\M{E}}_k\right]_{ij}\tr{\M{C}_k\M{G}_k\M{A}_{k-1}\M{G}_k^{\rm T}\M{C}_k^{\rm T}},
\end{aligned}
\end{equation}
where the third line follows from the fact that all the basis operators $\M{O}_i,~i=1\dotsc L$ are trace-nonincreasing operators, hence represent contractive transformations. Since unitary transformations preserve the trace, we further obtain
\begin{equation}
\begin{aligned}
\tr{\M{A}_k}&\le \tr{\M{A}_{k-1}}\bigg(1+\lambda_{\rm max}\{\M{C}_k^{\rm T}\M{C}_k\}\sum_{i=1}^L\sum_{j=1}^L  \left[\widetilde{\M{E}}_k\right]_{ij}\bigg).
\end{aligned}
\end{equation}
Note that
\begin{equation}
\begin{aligned}
\sum_{i=1}^L\sum_{j=1}^L \left[\widetilde{\M{E}}_k\right]_{ij} &\le \frac{1}{N_{\rm s}}\Big(\|\mathrm{vec}\{\M{P}_k\}\|_1 + \|\mathrm{vec}\{\V{p}_k\V{p}_k^{\rm T}\}\|_1\Big)\\
&=\frac{2}{N_{\rm s}},
\end{aligned}
\end{equation}
and that $\lambda_{\rm max}\{\M{C}_k^{\rm T}\M{C}_k\}\le 1$ since $\M{C}_k$ is a completely positive trace-preserving channel, hence is contractive. In light of this, the upper bound of $\tr{\M{A}_k}$ can now be simplified as follows:
\begin{equation}
\tr{\M{A}_k} \le \tr{\M{A}_{k-1}}\left(1+\frac{2}{N_{\rm s}}\right).
\end{equation}

From \eqref{initial_conditions} we have $\tr{\M{A}_0}= 1$ since $\V{v}_0$ is a unit vector, hence we obtain
\begin{equation}
\begin{aligned}
\tr{\M{A}_{N_{\rm G}}} &\le \prod_{k=1}^{N_{\rm G}} \left(1+\frac{2}{N_{\rm s}}\right)\\
&\le\exp\left(2N_{\rm G}N_{\rm s}^{-1}\right).
\end{aligned}
\end{equation}
Using \eqref{sigma_k}, we have
\begin{equation}
\begin{aligned}
\tr{\M{\Sigma}_{N_{\rm G}}} &= \tr{\M{A}_{N_{\rm G}}} - \tr{\V{v}_0\V{v}_0^{\rm T}} \\
&\le \exp\left(2N_{\rm G}N_{\rm s}^{-1}\right) - 1.
\end{aligned}
\end{equation}
Note that $\M{\Sigma_{N_{\rm G}}}$ is a positive semidefinite matrix, hence we have
\begin{equation}
\tr{\M{\Sigma_{N_{\rm G}}}} \ge \lambda_{\max}(\M{\Sigma_{N_{\rm G}}}),
\end{equation}
where $\lambda_{\max}(\cdot)$ denotes the maximum eigenvalue of a matrix. This implies that
\begin{equation}\label{bound_general}
\begin{aligned}
\sqrt{\mathbb{E}\{(\rv{r}-\widetilde{r})^2\}} &= \sqrt{\V{v}_{\rm ob}^{\rm T}\M{\Sigma}_{N_{\rm G}}\V{v}_{\rm ob}} \\
&\le \sqrt{\tr{\M{\Sigma_{N_{\rm G}}}}}\cdot \|\V{v}_{\rm ob}\| \\
&\le \sqrt{\exp\left(2N_{\rm G}N_{\rm s}^{-1}\right) - 1}\cdot \|\V{v}_{\rm ob}\|.
\end{aligned}
\end{equation}
Hence the proof is completed by applying \eqref{normalization}.
\end{IEEEproof}

Especially, for Pauli channels, we have the following simplified recursions:
\begin{equation}\label{recursion_A}
\begin{aligned}
\M{A}_k&=\mathbb{E}\{\widetilde{\RV{c}}_k\widetilde{\RV{c}}_k^{\rm T}\}\odot \M{G}_k\M{A}_{k-1}\M{G}_k^{\rm T},\\
\V{\mu}_k &= \M{G}_k\V{\mu}_{k-1}.
\end{aligned}
\end{equation}
In fact, we have
\begin{equation}\label{ckck}
\mathbb{E}\{\widetilde{\RV{c}}_k\widetilde{\RV{c}}_k^{\rm T}\} = \V{1}\V{1}^{\rm T} + \M{\Xi}_k,
\end{equation}
which follows from \eqref{residual_distribution}. Substituting \eqref{ckck} into \eqref{recursion_A}, we obtain
\begin{equation}\label{ak_recursion}
\M{A}_k = \M{G}_k\M{A}_{k-1}\M{G}_k^{\rm T} + \M{\Xi}_k\odot \M{G}_k\M{A}_{k-1}\M{G}_k^{\rm T}.
\end{equation}
Following the same line of reasoning as we used in the general case, we have
\begin{equation}\label{trace_ub_exact}
\begin{aligned}
\tr{\M{A}_k} &= \tr{\M{G}_k\M{A}_{k-1}\M{G}_k^{\rm T} + \M{\Xi}_k\odot \M{G}_k\M{A}_{k-1}\M{G}_k^{\rm T}} \\
&= \tr{\M{A}_{k-1}} + \tr{\M{\Xi}_k\odot \M{G}_k\M{A}_{k-1}\M{G}_k^{\rm T}} \\
&\le \tr{\M{A}_{k-1}} (1+\|\M{\Xi}_k\|_{\max}),
\end{aligned}
\end{equation}
the ``max norm'' $\|\cdot\|_{\max}$ is defined as
$$
\|\M{A}\|_{\max} := \max_{i,j} |[\M{A}]_{i,j}|.
$$
From \eqref{residual_distribution} we obtain
\begin{equation}\label{maxnorm}
\begin{aligned}
\|\M{\Xi}_k\|_{\max} &= \frac{1}{N_{\rm s}}\|\widetilde{\M{H}}\left(\M{P}_k-\V{p}_k\V{p}_k^{\rm T}\right)\widetilde{\M{H}}\odot \V{c}_k\V{c}_k^{\rm T}\|_{\max} \\
&\le \frac{1}{N_{\rm s}}\|\widetilde{\M{H}}\left(\M{P}_k-\V{p}_k\V{p}_k^{\rm T}\right)\widetilde{\M{H}}\|_{\max}\| \V{c}_k\V{c}_k^{\rm T}\|_{\max} \\
&\le \frac{1}{N_{\rm s}}\left(\|\widetilde{\M{H}}\M{P}_k\widetilde{\M{H}}\|_{\max}+\|\widetilde{\M{H}}\V{p}_k\V{p}_k^{\rm T}\widetilde{\M{H}}\|_{\max}\right),
\end{aligned}
\end{equation}
where the third line follows from the fact that $\V{c}_k$ represents a contractive transformation, so that $\V{c}_k\preceq \V{1}$. Note that every entry in $\widetilde{\M{H}}$ has an absolute value of $1$, and hence
\begin{equation}
\begin{aligned}
\|\widetilde{\M{H}}\M{P}_k\widetilde{\M{H}}\|_{\max} &\le \|\mathrm{vec}\{\M{P}_k\}\|_1 =1,
\end{aligned}
\end{equation}
and
\begin{equation}
\begin{aligned}
\|\widetilde{\M{H}}\V{p}_k\V{p}_k^{\rm T}\widetilde{\M{H}}\|_{\max} &\le \|\mathrm{vec}\{\V{p}_k\V{p}_k^{\rm T}\}\|_1 =1.
\end{aligned}
\end{equation}
Hence we arrive at exactly the same bound as given in \eqref{bound_general}.

\section{Proof of Proposition \ref{prop:qem_ger}}\label{sec:proof_qem_ger}
\begin{IEEEproof}
We start the proof from revisiting the inequality in \eqref{maxnorm}, and arrive at:
\begin{equation}
\begin{aligned}
\|\M{\Xi}_k\|_{\max} &\le \frac{1}{N_{\rm s}} \|\widetilde{\M{H}}\left(\M{P}_k-\V{p}_k\V{p}_k^{\rm T}\right)\widetilde{\M{H}}\|_{\max}\\
&\le \frac{1}{N_{\rm s}} \|\mathrm{vec}\{\M{P}_k-\V{p}_k\V{p}_k^{\rm T}\}\|_1.
\end{aligned}
\end{equation}

Next we construct an upper bound for the term $\|\mathrm{vec}\{\M{P}_k-\V{p}_k\V{p}_k^{\rm T}\}\|_1$. According to the sampling overhead of \ac{qem} in \cite{sof_analysis}, Assumption \ref{asu:fidelity} implies that
\begin{equation}
\|\V{\alpha}\|_1 \le \sqrt{1+\sigma_{\rm u}}.
\end{equation}
Since $\alpha_k^{(1)}\ge 1$, we have
\begin{equation}
\sum_{i\neq 1} |\alpha_k^{(i)}| \le \sqrt{1+\sigma_{\rm u}}-1.
\end{equation}
This further implies that
\begin{equation}\label{fidelity_constraints}
\begin{aligned}
p_k^{(1)} &\ge \frac{1}{\sqrt{1+\sigma_{\rm u}}}, \\
\sum_{i\neq 1} p_k^{(i)} &\le \sqrt{1+\sigma_{\rm u}}-1.
\end{aligned}
\end{equation}
Therefore, upon taking the entry-wise absolute value, we obtain
\begin{equation}
\left|\M{P}_k-\V{p}_k\V{p}_k^{\rm T}\right| \le \left(
                                                                    \begin{array}{cccc}
                                                                      \frac{\sigma_{\rm u}}{1+\sigma_{\rm u}} & p_k^{(2)} & \cdots & p_k^{(4^n)} \\
                                                                      p_k^{(2)} & p_k^{(2)} & p_k^{(2)}p_k^{(3)} & \cdots \\
                                                                      \vdots & p_k^{(3)}p_k^{(2)} & \ddots & \vdots \\
                                                                      p_k^{(4^n)} & \vdots & \cdots & p_k^{(4^n)} \\
                                                                    \end{array}
                                                                  \right).
\end{equation}
Here, the symbol ``$\le$'' stands for entry-wise ``not larger than''. Observe that summing up the first row, the first column and the main diagonal, by applying \eqref{fidelity_constraints}, we see that
\begin{equation}
\begin{aligned}
\|\mathrm{vec}\{\M{P}_k-\V{p}_k\V{p}_k^{\rm T}\}\|_1 &\le 3\left(\sqrt{1+\sigma_{\rm u}}-1\right) \\
&\hspace{3mm}+\frac{\sigma_{\rm u}}{1+\sigma_{\rm u}} + \|\mathrm{vec}\{\V{q}_k\V{q}_k^{\rm T}\}\|_1,
\end{aligned}
\end{equation}
where $\V{q}_k:=[p_k^{(2)}~\dotsc~p_k^{(4^q)}]^{\rm T} \in \mathbb{R}^{4^q-1}$. Note that
\begin{equation}
\begin{aligned}
\|\mathrm{vec}\{\V{q}_k\V{q}_k^{\rm T}\}\|_1 &= \V{1}^{\rm T}\V{q}_k\V{q}_k^{\rm T}\V{1}\\
&\le (\sqrt{1+\sigma_{\rm u}}-1)^2,
\end{aligned}
\end{equation}
implying that
\begin{equation}
\|\mathrm{vec}\{\M{P}_k-\V{p}_k\V{p}_k^{\rm T}\}\|_1\le \frac{5}{2}\sigma_{\rm u} + \frac{1}{4}\sigma_{\rm u}^2,
\end{equation}
which follows from that fact that
$$
\sqrt{1+x}-1 \le \frac{x}{2}
$$
holds for all $x\ge 0$. Hence we have
\begin{equation}
\|\M{\Xi}_k\|_{\max} \le\frac{1}{N_{\rm s}} \left(\frac{5}{2}\sigma_{\rm u} + \frac{1}{4}\sigma_{\rm u}^2\right),
\end{equation}
which proves \eqref{qem_rmse_bound2}. Thus the proof is completed.
\end{IEEEproof}

\bibliographystyle{ieeetran}
\bibliography{IEEEabrv,QEM}

\begin{thebibliography}{10}
\providecommand{\url}[1]{#1}
\csname url@samestyle\endcsname
\providecommand{\newblock}{\relax}
\providecommand{\bibinfo}[2]{#2}
\providecommand{\BIBentrySTDinterwordspacing}{\spaceskip=0pt\relax}
\providecommand{\BIBentryALTinterwordstretchfactor}{4}
\providecommand{\BIBentryALTinterwordspacing}{\spaceskip=\fontdimen2\font plus
\BIBentryALTinterwordstretchfactor\fontdimen3\font minus
  \fontdimen4\font\relax}
\providecommand{\BIBforeignlanguage}[2]{{%
\expandafter\ifx\csname l@#1\endcsname\relax
\typeout{** WARNING: IEEEtran.bst: No hyphenation pattern has been}%
\typeout{** loaded for the language `#1'. Using the pattern for}%
\typeout{** the default language instead.}%
\else
\language=\csname l@#1\endcsname
\fi
#2}}
\providecommand{\BIBdecl}{\relax}
\BIBdecl

\bibitem{nisq}
J.~Preskill, ``Quantum {C}omputing in the {NISQ} era and beyond,''
  \emph{{Quantum}}, vol.~2, pp. 1--21, Aug. 2018.

\bibitem{quantum_supremacy}
F.~Arute, K.~Arya, R.~Babbush \emph{et~al.}, ``{Quantum supremacy using a
  programmable superconducting processor},'' \emph{Nature}, vol. 574, no. 7779,
  pp. 505--510, Oct. 2019.

\bibitem{qecc}
A.~R. {Calderbank}, E.~M. {Rains}, P.~M. {Shor}, and N.~J.~A. {Sloane},
  ``Quantum error correction via codes over {GF$(4)$},'' \emph{IEEE Trans. Inf.
  Theory}, vol.~44, no.~4, pp. 1369--1387, Jul. 1998.

\bibitem{qit}
C.~H. {Bennett} and P.~W. {Shor}, ``Quantum information theory,'' \emph{IEEE
  Trans. Inf. Theory}, vol.~44, no.~6, pp. 2724--2742, Oct. 1998.

\bibitem{qecc_survey}
Z.~{Babar}, D.~{Chandra}, H.~V. {Nguyen}, P.~{Botsinis}, D.~{Alanis}, S.~X.
  {Ng}, and L.~{Hanzo}, ``Duality of quantum and classical error correction
  codes: {D}esign principles and examples,'' \emph{IEEE Commun. Surv. Tuts.},
  vol.~21, no.~1, pp. 970--1010, 1st quart. 2019.

\bibitem{qtecc}
D.~{Chandra}, Z.~{Babar}, H.~V. {Nguyen}, D.~{Alanis}, P.~{Botsinis}, S.~X.
  {Ng}, and L.~{Hanzo}, ``Quantum topological error correction codes: {T}he
  classical-to-quantum isomorphism perspective,'' \emph{IEEE Access}, vol.~6,
  pp. 13\,729--13\,757, 2018.

\bibitem{transversal}
E.~Knill, R.~Laflamme, and W.~H. Zurek, ``Resilient quantum computation: error
  models and thresholds,'' \emph{Proc. Roy.l Soc. London A, Math. Phys. Eng.
  Sci.}, vol. 454, no. 1969, p. 365–384, Jan. 1998.

\bibitem{threshold_thm}
D.~Aharonov and M.~Ben-Or, ``Fault-tolerant quantum computation with constant
  error rate,'' \emph{SIAM J. Comput.}, vol.~38, no.~4, p. 1207–1282, Jul.
  2008.

\bibitem{fault_tolerance}
D.~Gottesman, ``Theory of fault-tolerant quantum computation,'' \emph{Phys.
  Rev. A}, vol.~57, no.~1, p. 127, 1998.

\bibitem{shor}
P.~W. {Shor}, ``Algorithms for quantum computation: {D}iscrete logarithms and
  factoring,'' in \emph{Proc. 35th Annual Symp. Foundations of Computer
  Science}, Santa Fe, New Mexico, USA, Nov. 1994, pp. 124--134.

\bibitem{grover}
L.~K. Grover, ``A fast quantum mechanical algorithm for database search,'' in
  \emph{Proc. 28th Annual ACM Symp. Theory of Computing}, Philadelphia,
  Pennsylvania, USA, May 1996, pp. 212--219.

\bibitem{vqe}
P.~J. Love, J.~L. O'Brien, A.~Aspuru-Guzik, A.~Peruzzo, M.-h. Yung, X.-Q. Zhou,
  P.~Shadbolt, and J.~McClean, ``{A variational eigenvalue solver on a photonic
  quantum processor},'' \emph{Nature Commun.}, vol.~5, no.~1, pp. 1--7, Jul.
  2014.

\bibitem{vqe_theory}
J.~R. McClean, J.~Romero, R.~Babbush, and A.~Aspuru-Guzik, ``{The theory of
  variational hybrid quantum-classical algorithms},'' \emph{New Journal of
  Physics}, vol.~18, no.~2, pp. 1--22, Feb. 2016.

\bibitem{qaoa}
\BIBentryALTinterwordspacing
E.~Farhi, J.~Goldstone, and S.~Gutmann, ``A quantum approximate optimization
  algorithm,'' \emph{arXiv preprint}, 2014. [Online]. Available:
  \url{https://arxiv.org/abs/arXiv:1411.4028}
\BIBentrySTDinterwordspacing

\bibitem{vqlinear}
\BIBentryALTinterwordspacing
X.~Xu, J.~Sun, S.~Endo, Y.~Li, S.~C. Benjamin, and X.~Yuan, ``Variational
  algorithms for linear algebra,'' \emph{arXiv preprint}, 2019. [Online].
  Available: \url{https://arxiv.org/abs/1909.03898}
\BIBentrySTDinterwordspacing

\bibitem{sgd_vqa}
R.~Sweke, F.~Wilde, J.~J. Meyer, M.~Schuld, P.~K. F{\"a}hrmann,
  B.~Meynard-Piganeau, and J.~Eisert, ``Stochastic gradient descent for hybrid
  quantum-classical optimization,'' \emph{Quantum}, vol.~4, p. 314, 2020.

\bibitem{ansatz}
S.~Hadfield, Z.~Wang, B.~O’Gorman, E.~Rieffel, D.~Venturelli, and R.~Biswas,
  ``From the quantum approximate optimization algorithm to a quantum
  alternating operator ansatz,'' \emph{Algorithms}, vol.~12, no.~2, pp. 1--45,
  Feb. 2019.

\bibitem{performance_qaoa}
\BIBentryALTinterwordspacing
G.~E. Crooks, ``Performance of the quantum approximate optimization algorithm
  on the maximum cut problem,'' \emph{arXiv preprint}, 2018. [Online].
  Available: \url{https://arxiv.org/abs/1811.08419}
\BIBentrySTDinterwordspacing

\bibitem{barren_plateau}
J.~R. McClean, S.~Boixo, V.~N. Smelyanskiy, R.~Babbush, and H.~Neven, ``Barren
  plateaus in quantum neural network training landscapes,'' \emph{Nat.
  Commun.}, vol.~9, no.~1, pp. 1--6, Nov. 2018.

\bibitem{NIBP}
\BIBentryALTinterwordspacing
S.~Wang, E.~Fontana, M.~Cerezo, K.~Sharma, A.~Sone, L.~Cincio, and P.~J. Coles,
  ``Noise-induced barren plateaus in variational quantum algorithms,''
  \emph{arXiv preprint}, 2020. [Online]. Available:
  \url{https://arxiv.org/abs/2007.14384}
\BIBentrySTDinterwordspacing

\bibitem{practical_qem}
S.~Endo, S.~C. Benjamin, and Y.~Li, ``Practical quantum error mitigation for
  near-future applications,'' \emph{Phys. Rev. X}, vol.~8, no.~3, pp. 1--21,
  Jul. 2018.

\bibitem{sof_analysis}
Y.~Xiong, D.~Chandra, S.~X. Ng, and L.~Hanzo, ``Sampling overhead analysis of
  quantum error mitigation: {U}ncoded vs. coded systems,'' \emph{IEEE Access},
  vol.~8, pp. 228\,967--228\,991, Dec. 2020.

\bibitem{ryuji_cost_qem}
\BIBentryALTinterwordspacing
R.~Takagi, ``Optimal resource cost for error mitigation,'' \emph{Phys. Rev.
  Research}, vol.~3, p. 033178, Aug. 2021. [Online]. Available:
  \url{https://link.aps.org/doi/10.1103/PhysRevResearch.3.033178}
\BIBentrySTDinterwordspacing

\bibitem{vqe_nearterm}
N.~Moll, P.~Barkoutsos, L.~S. Bishop, J.~M. Chow, A.~Cross, D.~J. Egger,
  S.~Filipp, A.~Fuhrer, J.~M. Gambetta, M.~Ganzhorn \emph{et~al.}, ``Quantum
  optimization using variational algorithms on near-term quantum devices,''
  \emph{Quantum Science and Technology}, vol.~3, no.~3, 2018.

\bibitem{qaoa_decoding}
T.~{Matsumine}, T.~{Koike-Akino}, and Y.~{Wang}, ``Channel decoding with
  quantum approximate optimization algorithm,'' in \emph{Proc. IEEE Int. Symp.
  Inf. Theory (ISIT)}, Paris, France, Jul. 2019, pp. 2574--2578.

\bibitem{scalable_simulation}
P.~J. O’Malley, R.~Babbush, I.~D. Kivlichan, J.~Romero, J.~R. McClean,
  R.~Barends, J.~Kelly, P.~Roushan, A.~Tranter, N.~Ding \emph{et~al.},
  ``Scalable quantum simulation of molecular energies,'' \emph{Physical Review
  X}, vol.~6, no.~3, 2016.

\bibitem{bp2}
\BIBentryALTinterwordspacing
K.~Sharma, M.~Cerezo, L.~Cincio, and P.~J. Coles, ``Trainability of dissipative
  perceptron-based quantum neural networks,'' \emph{arXiv preprint}, 2020.
  [Online]. Available: \url{https://arxiv.org/abs/2005.12458}
\BIBentrySTDinterwordspacing

\bibitem{bp3}
\BIBentryALTinterwordspacing
M.~Cerezo, A.~Sone, T.~Volkoff, L.~Cincio, and P.~J. Coles,
  ``Cost-function-dependent barren plateaus in shallow quantum neural
  networks,'' \emph{arXiv preprint}, 2020. [Online]. Available:
  \url{https://arxiv.org/abs/2001.00550}
\BIBentrySTDinterwordspacing

\bibitem{resort1}
\BIBentryALTinterwordspacing
G.~Verdon, M.~Broughton, J.~R. McClean, K.~J. Sung, R.~Babbush, Z.~Jiang,
  H.~Neven, and M.~Mohseni, ``Learning to learn with quantum neural networks
  via classical neural networks,'' \emph{arXiv preprint}, 2019. [Online].
  Available: \url{https://arxiv.org/abs/1907.05415}
\BIBentrySTDinterwordspacing

\bibitem{resort2}
\BIBentryALTinterwordspacing
A.~Skolik, J.~R. McClean, M.~Mohseni, P.~van~der Smagt, and M.~Leib,
  ``Layerwise learning for quantum neural networks,'' \emph{arXiv preprint},
  2020. [Online]. Available: \url{https://arxiv.org/abs/2006.14904}
\BIBentrySTDinterwordspacing

\bibitem{qem}
K.~Temme, S.~Bravyi, and J.~M. Gambetta, ``Error mitigation for short-depth
  quantum circuits,'' \emph{Phys. Rev. Lett.}, vol. 119, no.~18, pp. 1--5, Nov.
  2017.

\bibitem{subspace_expansion1}
J.~I. Colless, V.~V. Ramasesh, D.~Dahlen, M.~S. Blok, M.~E. Kimchi-Schwartz,
  J.~R. McClean, J.~Carter, W.~A. de~Jong, and I.~Siddiqi, ``Computation of
  molecular spectra on a quantum processor with an error-resilient algorithm,''
  \emph{Phys. Rev. X}, vol.~8, Feb. 2018.

\bibitem{subspace_expansion2}
J.~R. McClean, Z.~Jiang, N.~C. Rubin, R.~Babbush, and H.~Neven, ``Decoding
  quantum errors with subspace expansions,'' \emph{Nat. Commun.}, vol.~11,
  no.~1, pp. 1--9, 2020.

\bibitem{symmetry_verification}
X.~Bonet-Monroig, R.~Sagastizabal, M.~Singh, and T.~E. O'Brien, ``Low-cost
  error mitigation by symmetry verification,'' \emph{Phys. Rev. A}, vol.~98,
  Dec. 2018.

\bibitem{qem_exp}
C.~Song, J.~Cui, H.~Wang, J.~Hao, H.~Feng, and Y.~Li, ``Quantum computation
  with universal error mitigation on a superconducting quantum processor,''
  \emph{Science Advances}, vol.~5, no.~9, 2019.

\bibitem{ncbook}
M.~A. Nielsen and I.~L. Chuang, \emph{Quantum Computation and Quantum
  Information}, 2nd~ed.\hskip 1em plus 0.5em minus 0.4em\relax New York, NY,
  USA: Cambridge University Press, 2011.

\bibitem{ptm}
J.~M. Chow, J.~M. Gambetta, A.~D. C\'orcoles, S.~T. Merkel, J.~A. Smolin,
  C.~Rigetti, S.~Poletto, G.~A. Keefe, M.~B. Rothwell, J.~R. Rozen, M.~B.
  Ketchen, and M.~Steffen, ``Universal quantum gate set approaching
  fault-tolerant thresholds with superconducting qubits,'' \emph{Phys. Rev.
  Lett.}, vol. 109, no.~6, pp. 1--5, Aug. 2012.

\bibitem{pauli_channel}
M.~A. Cirone, A.~Delgado, D.~G. Fischer, M.~Freyberger, H.~Mack, and
  M.~Mussinger, ``Estimation of quantum channels with finite resources,''
  \emph{Quantum Information Processing}, vol.~1, no.~5, pp. 303--326, Oct.
  2002.

\bibitem{gst_intro}
\BIBentryALTinterwordspacing
D.~Greenbaum, ``Introduction to quantum gate set tomography,'' \emph{arXiv
  preprint}, 2015. [Online]. Available: \url{https://arxiv.org/abs/1509.02921}
\BIBentrySTDinterwordspacing

\bibitem{adiabatic}
E.~Farhi, J.~Goldstone, S.~Gutmann, J.~Lapan, A.~Lundgren, and D.~Preda, ``A
  quantum adiabatic evolution algorithm applied to random instances of an
  {NP}-complete problem,'' \emph{Science}, vol. 292, no. 5516, pp. 472--475,
  2001.

\bibitem{mud}
S.~Verdu \emph{et~al.}, \emph{Multiuser detection}.\hskip 1em plus 0.5em minus
  0.4em\relax Cambridge university press, 1998.

\bibitem{quantum_mud}
P.~{Botsinis}, S.~X. {Ng}, and L.~{Hanzo}, ``Fixed-complexity quantum-assisted
  multi-user detection for {CDMA} and {SDMA},'' \emph{IEEE Trans. Commun.},
  vol.~62, no.~3, pp. 990--1000, Mar. 2014.

\end{thebibliography}

\begin{IEEEbiography}[{\includegraphics[width=1in,height=1.25in,clip,keepaspectratio]{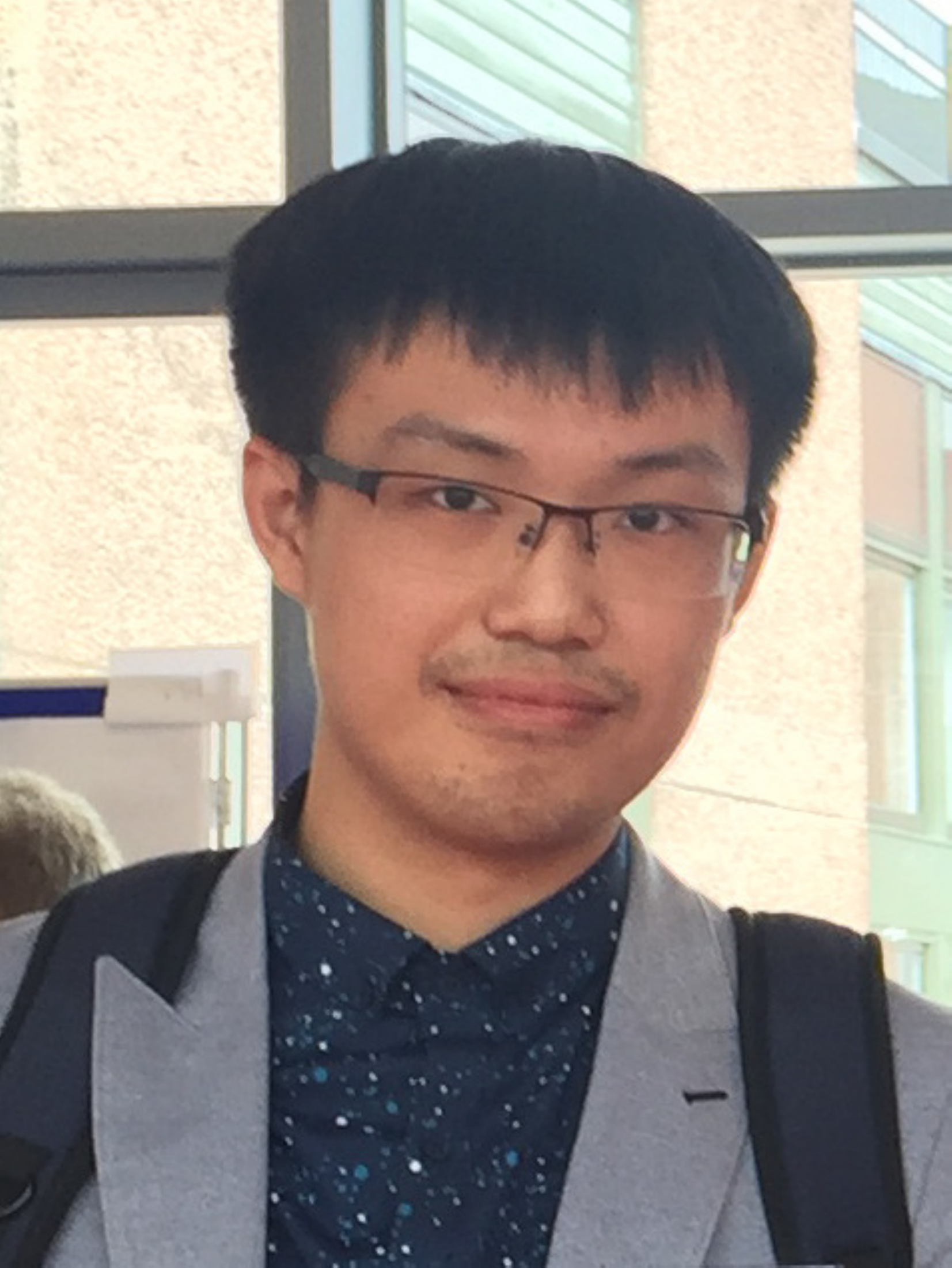}}]{\bf Yifeng Xiong} received his B.S. degree in information engineering, and the M.S. degree in information and communication engineering from Beijing Institute of Technology (BIT), Beijing, China, in 2015 and 2018, respectively. He is currently pursuing the PhD degree with Next-Generation Wireless within the School of Electronics and Computer Science, University of Southampton. His research interests include quantum computation, quantum information theory, graph signal processing, and statistical inference over networks.
\end{IEEEbiography}

\begin{IEEEbiography}[{\includegraphics[width=1in,height=1.25in,clip,keepaspectratio]{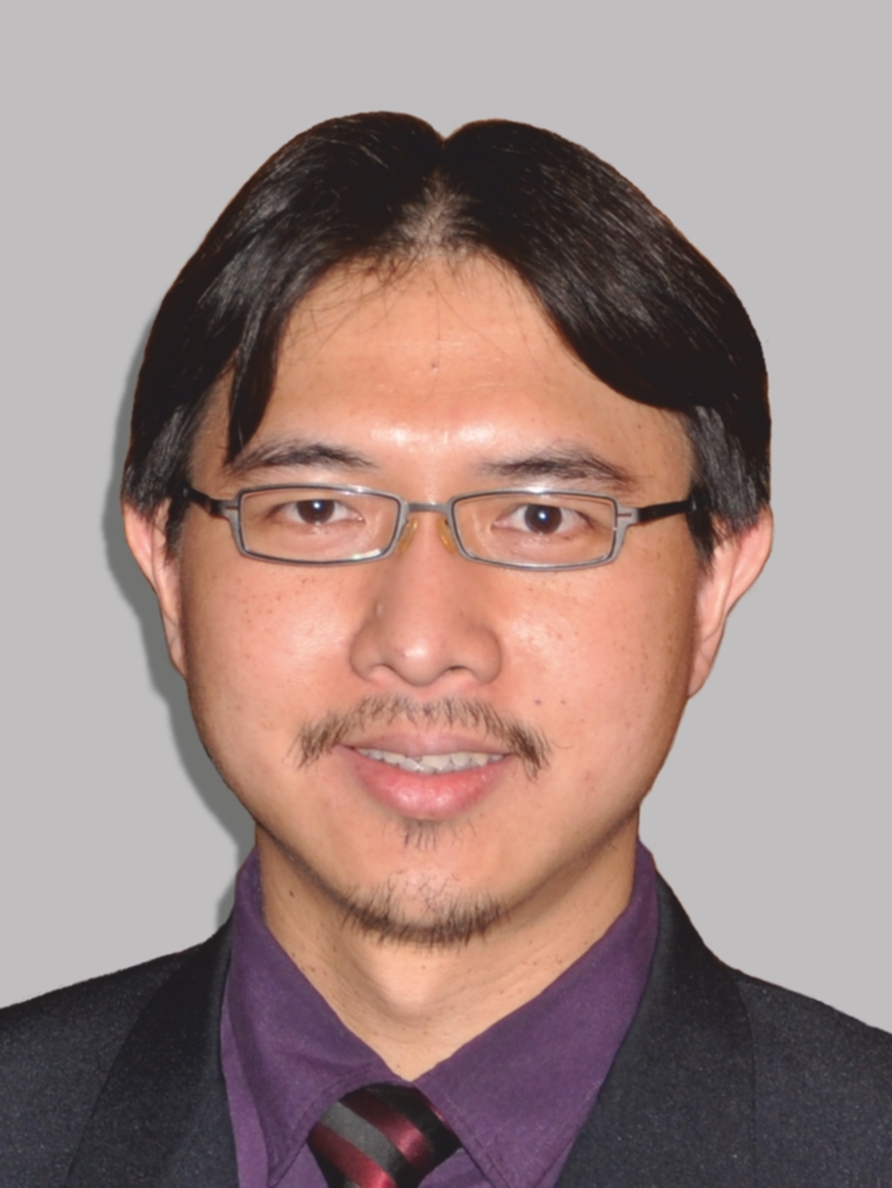}}]{\bf Soon Xin Ng} (S'99-M'03-SM'08) received the B.Eng. degree (First class) in electronic engineering and the Ph.D. degree in telecommunications from the University of Southampton, Southampton, U.K., in 1999 and 2002, respectively. From 2003 to 2006, he was a postdoctoral research fellow working on collaborative European research projects known as SCOUT, NEWCOM and PHOENIX. Since August 2006, he has been a member of academic staff in the School of Electronics and Computer Science, University of Southampton. He was involved in the OPTIMIX and CONCERTO European projects as well as the IU-ATC and UC4G projects. He was the principal investigator of an EPSRC project on ``Cooperative Classical and Quantum Communications Systems''. He is currently a Professor of Next Generation Communications at the University of Southampton.

His research interests include adaptive coded modulation, coded modulation, channel coding, space-time coding, joint source and channel coding, iterative detection, OFDM, MIMO, cooperative communications, distributed coding, quantum communications, quantum error correction codes, joint wireless-and-optical-fibre communications, game theory, artificial intelligence and machine learning. He has published over 260 papers and co-authored two John Wiley/IEEE Press books in this field.

He is a Senior Member of the IEEE, a Fellow of the Higher Education Academy in the UK, a Chartered Engineer and a Fellow of the IET. He acted as TPC/track/workshop chairs for various conferences. He serves as an editor of Quantum Engineering. He was a guest editor for the special issues in IEEE Journal on Selected Areas in Communication as well as editors in the IEEE Access and the KSII Transactions on Internet and Information Systems. He is one of the Founders and Officers of the IEEE Quantum Communications \& Information Technology Emerging Technical Subcommittee (QCIT-ETC).
\end{IEEEbiography}

\begin{IEEEbiography}
[{\includegraphics[width=1in,height=1.25in,clip,keepaspectratio]{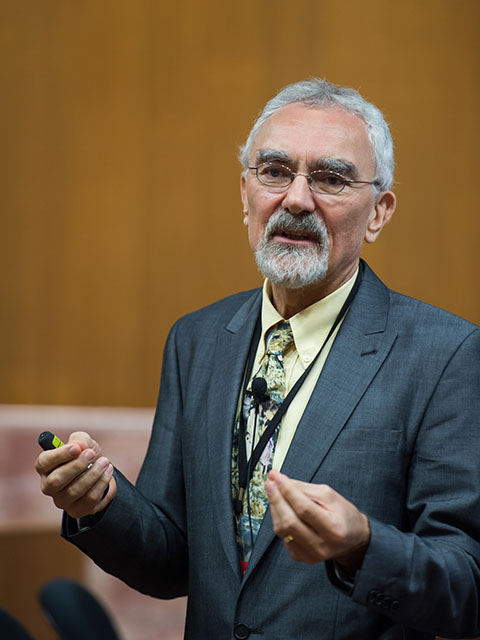}}]{\bf Lajos Hanzo} (http://www-mobile.ecs.soton.ac.uk, https://en.wikipedia.org/wiki/Lajos\_Hanzo) (FIEEE’04) received his Master degree and Doctorate in 1976 and 1983, respectively from the Technical University (TU) of Budapest. He was also awarded the Doctor of Sciences (DSc) degree by the University of Southampton (2004) and Honorary Doctorates by the TU of Budapest (2009) and by the University of Edinburgh (2015).  He is a Foreign Member of the Hungarian Academy of Sciences and a former Editor-in-Chief of the IEEE Press.  He has served several terms as Governor of both IEEE ComSoc and of VTS.  He has published 2000+ contributions at IEEE Xplore, 19 Wiley-IEEE Press books and has helped the fast-track career of 123 PhD students. Over 40 of them are Professors at various stages of their careers in academia and many of them are leading scientists in the wireless industry. He is also a Fellow of the Royal Academy of Engineering (FREng), of the IET and of EURASIP. He is the recipient of the 2022 Eric Sumner Field Award.
\end{IEEEbiography}

\end{document}